\documentclass[amsmath, amssymb,aps, pre,twocolumn,10pt]{revtex4-1}

\usepackage{url}
\usepackage{amsmath}
\usepackage{amssymb}

\usepackage{makeidx}         
\usepackage{graphicx}        
\usepackage[bottom]{footmisc}

\makeindex             

\usepackage[margin=1.99cm]{geometry}

\renewcommand{\vec}[1]{\boldsymbol{#1}}

\usepackage{ifthen}
\newcommand{\eq}[2][]{
  \ifthenelse{\equal{#1}{}}{%
    \begin{equation*}%
     #2%
     \end{equation*}%
   }{%
     \begin{equation}%
       \label{eq:#1}%
       #2%
     \end{equation}%
   }%
}
\newcommand{\meq}[2][]{
   \ifthenelse{\equal{#1}{}}{%
     \begin{equation*}%
       \begin{split}%
        #2%
       \end{split}%
     \end{equation*}%
   }{%
    \begin{equation}%
    \label{eq:#1}%
      \begin{split}%
        #2%
      \end{split}%
    \end{equation}%
   }%
}

\renewcommand{\eqref}[1]{(\protect\ref{eq:#1})}

\begin{document}

\title{Phase Transitions of Cellular Automata}
\author{Franco Bagnoli}
\affiliation{Dipartimento di Fisica ed Astronomia and CSDC,  Universit\`a degli Studi di Firenze, via G. Sansone 1, 50019 Sesto Fiorentino, Italy. Also INFN, Sez. di Firenze.}
\email{franco.bagnoli@unifi.it}

\author{Ra\'ul Rechtman}
\affiliation{Instituto de Energ\'i{}as Renovables, Universidad Nacional 
 Aut\'onoma de M\'exico, Apdo.\ Postal 34, 62580 Temixco Mor.,  Mexico}
 \email{rrs@ier.unam.mx}
%
%

\begin{abstract}
We explore some aspects of phase transitions in cellular automata. We start recalling the standard formulation of statistical mechanics of discrete systems (Ising model), illustrating the Monte Carlo approach as Markov chains and stochastic processes. We then formulate the cellular automaton problem using simple models, and illustrate different types of possible phase transitions: density phase transitions of first and second order, damage spreading, dilution of deterministic rules, asynchronism-induced transitions, synchronization phenomena, chaotic phase transitions and the influence of the topology. We illustrate the improved mean-field techniques and the phenomenological renormalization group approach. 
\end{abstract}

\maketitle

\section{Introduction. Equilibrium systems}
\label{sec:equilibrium}

Traditionally, phase transitions have been associated to the equilibrium state of stochastic systems. The main idea is the following~\cite{Encyclopaedia}. The ``fundamental'' description of a physical system is based on quantum mechanics, but let us suppose that classical physics is sufficiently accurate. Thus, our system under study is composed by a large number of degrees of freedom, whose evolution is given by an equivalently large set of  deterministic differential equations. We can study small portions of such systems using a molecular dynamics approach. In general, we are not interested in the characteristics of single trajectories, but rather on the average properties of systems with different initial conditions (or boundary terms): the statistical ensembles.

In many cases~\footnote{Or: in those cases that have been studied at depth.} the modes associated to the degrees of freedom can be divided into fast and slow ones. The fast ones can be approximated by ``noise'' and, upon equilibrium, be termed as ``temperature''. The evolution of the slow degrees of freedom (the dynamics)  becomes thus stochastic. 
By considering different initial conditions and different realizations of the noise, we can study the ensemble of stochastic trajectories , i.e.,  the temporal evolution of a probability distribution of possible states. Let us suppose that the evolution of the system can be approximated by a Markovian process. 

In closed systems, supposing that all degrees of freedom are coupled (and that the dynamics is chaotic), we can invoke ergodicity (at least for a large part of the phase space, in the presence of phase transitions and ergodicity breaking) and equipartition. Looking at a small portion of a closed system (the canonical ensemble), we can thus define the temperature and the thermal equilibrium.

The ergodicity is related to the accessible  volume of the phase space and therefore to the Boltzmann entropy, defined as the logarithm of this accessible volume. In continuous systems we have to introduce a coarse graining (invoking quantum mechanics) of the space, in order to reduce the volume to a number. 

In this context, a phase transition is related to a variation of the accessible volume (and this to the entropy), for instance triggered by the temperature (that here becomes our control parameter). For high temperatures, the whole space is accessible. For a given system size, and below a critical temperature, the  phase space is partitioned into ``valleys'' that tend to trap the dynamics for a long period. The presence of these valleys is put into evidence by choosing one or more suitable observables, function of the slow degrees of freedom: the order parameter(s).  

Supposing that the characteristic permanence time inside a valley grows (exponentially) on the size of the system, in the infinite-size limit we have a true partition of the phase space. According with the initial condition, the dynamics ends into one of the available attractors (in the case of the deterministic, high-dimensional equations of motion) or, in the language of Markov processes, we can say that the transition matrix is no more irreducible or that the largest eigenvalue (of value one) is degenerate. We can use the same word, ``attractor'' also to characterize the stochastic convergence to different asymptotic probability distributions. 

The principle of least information or maximum entropy allows to obtain the equilibrium (asymptotic) probability distribution of a closed system, the flat distribution over the available phase space, which is consistent with the ergodic hypothesis. In the canonical ensemble, this leads to the Boltzmann distribution 
\eq{
  P(\vec{x}) = \dfrac{1}{\mathcal{Z}}\exp\left(-\beta \mathcal{H}(\vec{x})\right),
}
where $\beta = 1/T$ is the inverse temperature (in the unit of energy), $\vec{x}$ is the set of slow degrees of freedom, and $\mathcal{H}(\vec{x})$ is the related energy (Hamiltonian). 
The quantity $Z$ (the partition function) is the normalization constant 
\eq{ 
  Z\equiv Z(T,V,N) = \sum_{\vec{x}} \exp\left(-\beta \mathcal{H}(\vec{x})\right).
}
By identifying the previous Boltzmann entropy with the thermodynamic one, the partition function is linked to the Helmotz free energy $F = - T \log(Z)$ and by knowing $Z$ one can obtain the thermodynamic properties of the system. Unfortunately, computing $Z$ is hard task. 

The partition function owes its name to the fact that if the Hamiltonian can be written as a sum of two parts that depend on different variables, $\vec{x} = (\vec{y}, \vec{z})$, $ \mathcal{H}(\vec{x}) =  \mathcal{H}(\vec{y}, \vec{z}) =  \mathcal{H}_y(\vec{y})+ \mathcal{H}_z(\vec{z})$ and due to the properties of the exponential function, $Z=Z_yZ_z$ and $F=F_x+F_z$, consistently with the extensive property of the free energy.  Notice that in reality we still need a small interaction term $\mathcal{H}_{yz}(\vec{y}, \vec{z})$ in order to have the thermal equilibrium (same temperature) between the $y$ and $z$ subsystems. 

Since the Hamiltonian is, for many systems, given by a sum of separate kinetic energies plus a non-separable configurational energy, it is possible in this way to study the kinetic parts separately (they constitute simple one-variable problems). One is thus left with a difficult configurational problem  with a smaller number of degrees of freedom. 

\begin{figure}[t]
\begin{center}
\includegraphics[width=0.82\columnwidth,bb=50 50 450 320]{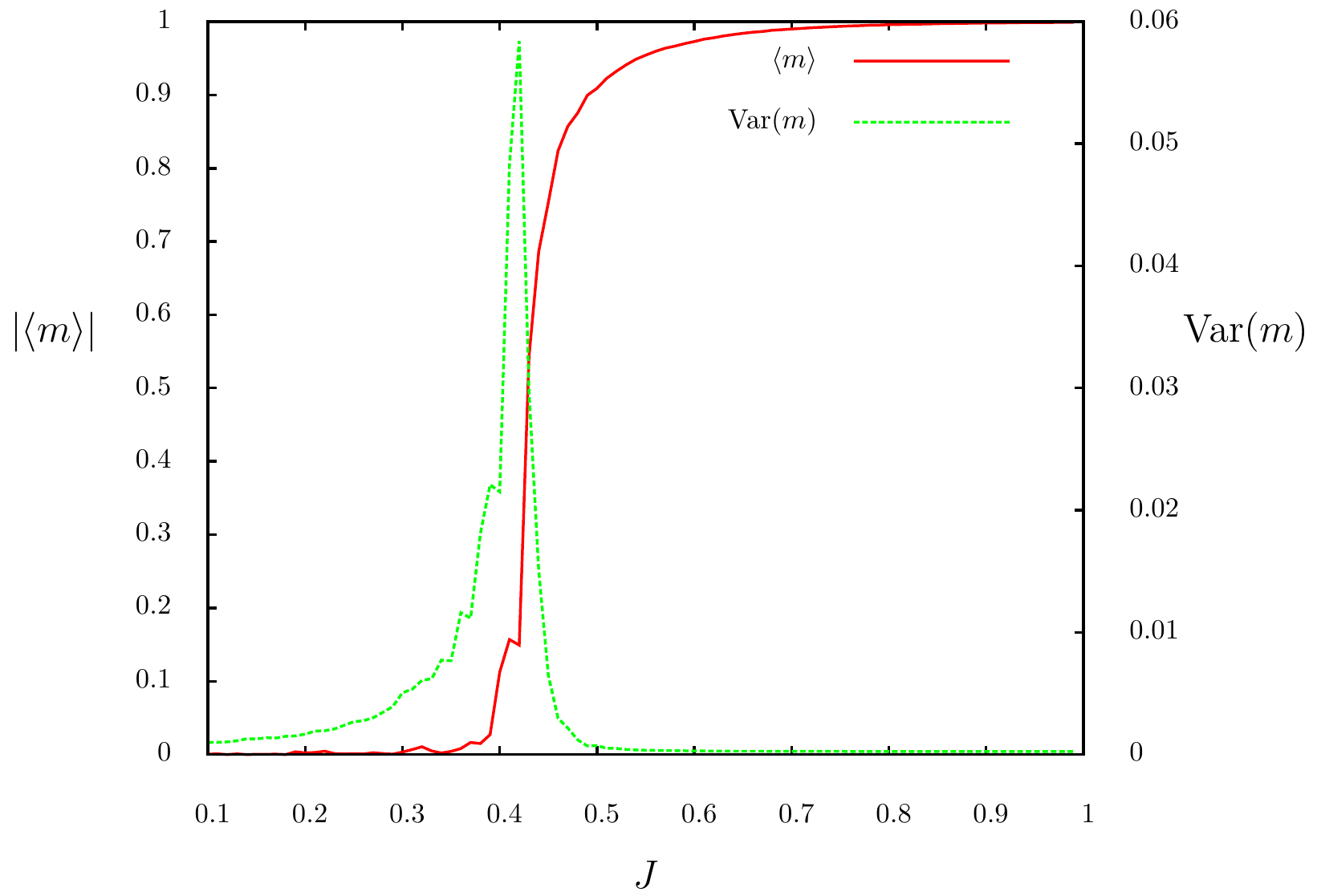} 
\caption{\label{fig:ising} Phase transition for the Ising model in 2D with nearest neighbour interactions. Average magnetization $|\langle m\rangle|$ and variance as a function of the rescaled coupling $J$ for $H=0$. Size $40\times40$, $T=4000$, transient $4\cdot 10^4$.}
\end{center}
\end{figure}

\subsection{Topology}
Since we are speaking of configurations, let us introduce some notation.  We always represent the configuration using only one index, like $\vec{s} = (s_1, s_2, \dots, s_N)$. The connections among sites are defined by the adjacency matrix $a_{ij}$ which takes value 1 is $i$ and $j$ are connected, and zero otherwise (one can obviously also introduce the strength of the interactions. For defining lattices, it is convenient to introduce also the truth function $[\cdot]$, which takes value 1 if $\cdot$ is true, and zero otherwise (similar to the Kronecker delta, but more flexible), and the Boolean functions OR ($\vee$), AND (multiplication) and XOR ($\oplus$). 
 
We can start from regular lattices of linear dimension $L$, for instance in 1D with nearest-neighbour interactions,  we have $N=L$ and  
$a_{ij} = [j = (i\pm 1) \mod L]$ (for periodic boundary conditions). In 2D $N=L^2$ and  $a_{ij} = [j =( i\pm1) \mod L] \vee [j = (i \pm L) \mod L^2]$, and so on. A random network (Erd\H{o}s R\'e{}nyi)  can be expressed as $a_{ij} = [r_{ij} < p]$ where $r_{ij}$ is a random number extracted from a flat distribution $0\le r_{ij}< 1$; the indices $i$ and $j$ indicate that we have to extract a random number for each link. In this way we generate a non-symmetric matrix, for having it symmetric (without self-interactions) it is sufficient to generate in this way only the part $i>j$, and copy the rest $a_{ij} = [r_{ij} < p][i>j] \vee a_{ji} [i<j]$.  

The connectivity $k_i$ of a site $i$ is the number of link connecting site $i$ to other sites (for non-symmetric matrices one has in- and out-connectivities): $k_i = sum_j a_{ij}$. For regular lattices the connectivity is constant,  while for random lattices defined above, if $k\ll N$, the connectivity is distributed as a Gaussian 
\eq{
 P(k) = \binom{N-1}{p} p^k (1-p)^{N-1-k} \simeq \dfrac{(Np)^k e^{-Np}}{k!}
 }
 for large $N$ and $Np=\text{const}$.
 
The neighbourhood of a site $i$ is the set of connected sites: $(s_j)_{a_{ij}=1}$ (for symmetric networks).

\subsection{Ising model}

We can now introduce the famous Ising model, that is a pure configurational model (one can think that the kinetic energy has been already partitioned). 
 Given an adiacency matrix $a_{ij}$, a coupling $J$ and a magnetic field $H$, the energy $\mathcal{H}(\vec{s})$ of a spin configuration $\vec{s} = (s_1, s_2, \dots, s_N)$ ($s_i=\pm 1$) is given by 

\eq[ising]{
  \mathcal{H}(\vec{x}) = - J \sum_{i,j} a_{ij} s_i s_j - H \sum_i s_i.
}

The Ising probability distribution is 
\eq{
  P(\vec{s}) = \dfrac{1}{Z} \exp\left(\beta  \left(J \sum_{i,j} a_{ij} s_i s_j + H \sum_i s_i\right)\right),
}
and we can absorb the inverse temperature $\beta$ in the parameters $J$ and $H$ (control parameters).

The magnetization $m$ is defined as 
\eq{
  m = m(J,H)= \sum_{\vec{s}} \left(P(\vec{s}) \frac{1}{N}\sum_i s_i \right).
}
It constitutes a suitable observable for this problem, as also its variance. From Onsager solution in 2D and zero magnetic field~\cite{Onsager}, we should observe a phase transition at $J_c \simeq 0.44$, with a transition from $m=0$ to $m\neq 0$ and the divergence of its variance, see Fig.~\ref{fig:ising} for a numerical simulation.

\subsection{Monte Carlo}

In all soluble systems in statistical mechanics (including the two-dimensional zero-field Ising), a transformation of variables (for instance the Fourier representation) or an approximation (like Einstein's solid or the mean-field one) it has been found, making the system separable. For non-separable problems we have to resort to Monte Carlo computations. 

In some sense, Monte Carlo computations represent the backward direction of the path followed so far. We look for a stochastic process that has  the equilibrium distribution $P^\text{eq}(\vec{x})$ as the asymptotic one.  In so doing, we are free to choose the dynamics, preferably looking for those for which the observables quickly converge to their asymptotic values. 

The ``golden rule'' is that of the detailed balance principle (which is only  a sufficient condition). Given any two configurations $\vec{x}$ and $\vec{y}$, we just need to choose the corresponding transition probabilities of a Markov process such that

\eq{ 
  \dfrac{M(\vec{y}|\vec{x})}{M(\vec{x}|\vec{y})} = \dfrac{P^\text{eq}(\vec{y})}{P^\text{eq}(\vec{x})},
}
where $M(x|y)$ is the conditional probability of getting $x$ given $y$, with $\sum_x M(x|y)=1$.

The problem of efficiency is that of choosing $\vec{y}$ giving $\vec{x}$ such that the exploration is ergodic, and we have a fast convergence of observables to their asymptotic values. In order to have that, the energies of the two configurations cannot be too different, and in particular we cannot draw $\vec{y}$ at random. Rather, one generally chooses $\vec{y}=\vec{x}$ except for a random site $i$. For instance, in the Ising model we can have $y_i=s'_i=-x_i = -s_i$.  The transition probability is 

\eq{
  \dfrac{M(\vec{s}'|\vec{s})}{M(\vec{s}|\vec{s}')} = \exp(-\beta \mathcal{H}(\vec{s}') - \mathcal{H}(\vec{s}).
}
Inserting Eq.~\eqref{ising} and simplifying,
\eq{
  \dfrac{M(\vec{y}|\vec{x})}{M(\vec{x}|\vec{y})} = \exp\bigl(-\beta s_i (H + J \sum_j a_{ij} s_j)\bigr).
}

There are many possible recipes, the one that we examine is the \emph{heat bath} dynamics, for which the probability that spin $i$ takes value $s'_i$  is 
\meq[isingprob]{
  &\tau(s'_i|(s_j)_{a_{ij}=1}) \\
  &= \dfrac{\exp(  s_i (H + J \sum_j a_{ij} s_j)}{\exp(\beta s_i  (H + J \sum_j a_{ij} s_j)) + \exp(- s_i (H + J \sum_j a_{ij} s_j))} \\
  &=   \dfrac{1}{1+  \exp(-2 s_i (H + J \sum_j a_{ij} s_j))}.
}
In practice, the Markov entry $M(\vec{s}'|\vec{s} = \tau(s'_i|(s_j)_{a_{ij}=1})$, all other spins remaining the same. 

\subsection{Monte Carlo trajectories}
We have described so far the  behaviour of the system from a probabilistic point of view, but in practice we have to compute a \textit{stochastic trajectory} that depends on a certain number of random numbers. 

The algorithm follow these steps, for each time $t$:
\begin{enumerate}
\item Choose a site at random $i = \lfloor N r(t)\rfloor$, $0\le r_1(t) < 1$ is a random number from a uniform distribution, and $ \lfloor x\rfloor$ represents the largest integer smaller than $x$. 
\item Compute $\Delta E =  - s_i (H + J \sum_j a_{ij} s_j)$ and $p = 1/(1+\exp(2\Delta E))$. 
\item The new value of site $i$ is $s'_i = \operatorname{sign}(p - r_2(t))$. 
\end{enumerate}

So we have a trajectory that depends on the random numbers $r_1(t)$ and $r_2(t)$. The time dependence of these numbers means that they change whenever $t$ does. Actually, we can also think of extracting the whole series of random numbers $(r(t'))_{t'=1,\dots,t}$ at beginning, and then perform the simulation using this set of numbers. In this way, they behave as a \textit{stochastic field} that changes at every time, and thus it is evident that the configuration at time $t$ is a function of the initial configuration and of the random field 
\eq{
  \vec{s}(t) = \vec{s}(\vec{s}(0), (r(t'))_{t=1,\dots,t})).
}
Notice that we can represent the evolution of the system on a time-space lattice, in which a site $i$ at time $t$ is located in the node of coordinates $(i, t)$ (denoted as transversal and parallel directions with respect to time). Each site is connected to its neighbourhood (for computing $\Delta E$ and to itself (for retrieving its old value). The random numbers can be distributed on this lattice. In this view, the stochastic trajectories become deterministic once that we have chosen the stochastic field. 

In principle, we should derive the probability distribution by averaging over the initial configuration $\vec{s}(0)$ and the stochastic field $ (r(t'))_{t=1,\dots,t}$, according with the statistical ensemble we are interested in. In practice, in many cases (but not always) for the computation of observables it is sufficient to perform just one simulation, averaging over time (after a transient), provided that the time is large enough. The property that time averages are equivalent to ensemble averages is called ergodicity.

The results of a simulation for the Ising model are shown in Fig.~\ref{fig:ising}.

In some cases, the system contains other quenched disorder, for instance if the couplings $J_{ij}$ or the external field $H_i$ depend on site indexes. Again, in principle one should average over the realization of disorder, but in many cases (not always), it is sufficient to perform a single simulation for a large enough system.  This is called the \textit{self-averaging} property and in somehow analogous to ergodicity. We shall see that these properties break at phase transitions.

\subsection{Equilibrium phase transitions}
There is a vast literature about phase transition in equilibrium statistical physics. We want here just recall some property that can be useful for extending the concept to arbitrary systems, not necessarily in equilibrium. 

Phase transitions are characterized by a change of the value of some observable in correspondence of a precise value of a control parameter, say the magnetization $m(J, H)$.
 In practice we can say that the dynamics of the system changes its structure in correspondence of a phase transition, for instance the phase space may effectively break in two zones that do not communicate at all. This is equivalent to say that the system is no more ergodic, and we speak of \textit{ergodicity breaking}.

However, we have a kind of contradiction here: we chose the Monte Carlo dynamics to be ergodic (i.e., there is a finite probability to go from any configuration to any other one), so how can ergodicity breaking occur? Actually, this breaking only manifests itself in a limiting procedure: for a finite system (finite $N$), and long enough time, all the phase space is visited (it is finite), and therefore the average of observables take a unique value. However, near  the phase transition,  the observables (say, the magnetization in the Ising model) maintain the same value for long periods, with occasional switches from one extreme to another. So, while its average value has a certain value (say, zero), one never observes such value! The time that the system spends on one phase become longer  as we approach the critical value of the control parameter and (exponentially) as we increase the system size.  

If we take first the limit of infinite system size and then that of infinite time, we observe the ergodicity breaking. In practice, it is sufficient to use a large enough system. In the language of stochastic trajectories, there are two low-energy valley separated by a high (energy) and/or large (entropy) barrier. in order to connect the two valleys, a path should climb the separating saddle,  and the associated probability becomes smaller and smaller with the system size, in the vicinity of the phase transition and above. 

In the language of Markov processes, we always have an irreducible transition matrix (since the dynamics is ergodic), but in the previous limit the time-product of matrices (denoted as $\vec{M}$) effectively breaks in two (or more) sub-matrices, that do not communicate
\eq{
\vec{M} =\begin{pmatrix}
\vec{M}_1 & \epsilon \\
\epsilon & \vec{M}_2
\end{pmatrix} \xrightarrow{N\rightarrow \infty} \begin{pmatrix}
\vec{M}_1 & 0 \\
0 & \vec{M}_2
\end{pmatrix},
}
where the $\epsilon$ denote the paths that connects the two valleys. The asymptotic distribution $P^\text{eq}(\vec{x})$ is proportional to the eigenvector of $M$ with eigenvalue 1. At phase transition this eigenvalue becomes degenerate and we have two or more asymptotic distributions, with different ``basins''. 

We can introduce the correlation function
\meq{
  C(\rho, \tau) =& \left(\sum_{i=1}^N \sum_{t=1}^T x_i(t) s_{i+\rho}(t+\tau)\right) -\\
  &  \left(\sum_{i=1}^N \sum_{t=1}^T s_i(t)\right)\left(\sum_{i=1}^N \sum_{t=1}^T s_{i+\rho}(t+\tau)\right)
}
The observables can be defined in terms of the correlation function. 

The correlation function is expected to decrease exponentially
\eq{
 C (\rho, \tau) \sim \exp\left(-\dfrac{\rho}{\xi_\perp}\right)\exp\left(-\dfrac{\tau}{\xi_\parallel}\right). 
}
defining the correlation length $\xi_\perp$ and $\xi_\parallel$ (with respect to time). 
 
At a phase transition (non-analytical behaviour of some  observables like discontinuities, divergence or angular points) the correlation lengths can stay finite (first-order phase transitions) or diverge (second-order phase transitions). In the latter case, 
\eq{
  \xi(J,H; N) \sim N^\alpha \tilde \xi\left( \dfrac{J}{N^\gamma},\frac{H}{N^\delta} \right),
}  
where $\alpha, \gamma, \delta$ are critical exponents. Also observables like the magnetization exhibit similar scaling behaviour.   This phenomenology extends to systems defined directly by stochastic transition probabilities. 

\begin{figure}[t]
\begin{center}
\includegraphics[width=1\columnwidth]{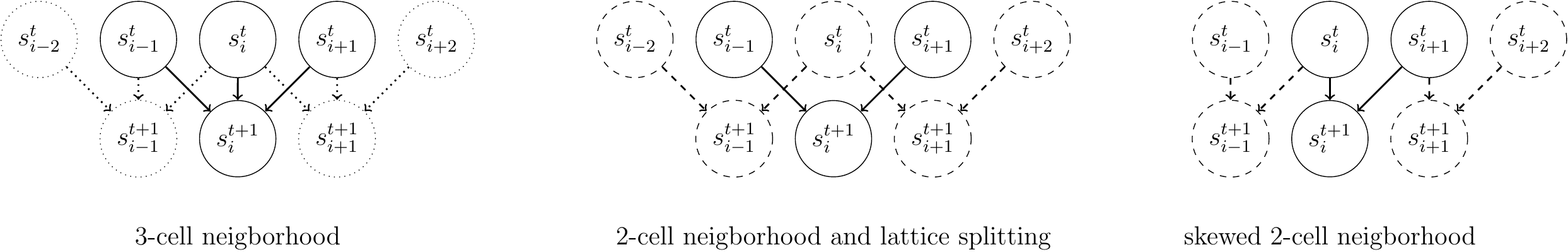}
\caption{\label{fig:neigh} Different cellular automata neighbourhood.}
\end{center}
\end{figure}

\section{Probabilistic Cellular Automata}
\label{sec:PCA}

The path that we have illustrated so far cannot be applied to all problems: in many cases we are looking for the asymptotic properties of a system that is just defined in terms of the local transition probabilities, of which  Probabilistic Cellular Automata (PCA) are prototypical examples. 

Cellular automata are defined as time-dependent models, so one should consider a space-time lattice with asymmetric connections at least in the time direction. For a site $i$ at time $t$, its neighbourhood is a set of other sites at previous time, see for instance Fig.~\ref{fig:neigh}. 

Probabilistic Cellular Automata are Markov chains for which the matrix elements are given by product of local transition probabilities (generally uniform). 

Let us denote by $\vec{x}=(x_0, x_1, \dots, x_{N-1})$ a possible  configuration of the system ($x_i$  is discrete, for instance $x_i=0,1$).  The state of the statistical ensemble at time $t$ is expressed by a probability distribution $P(\vec{x}, t)$, whose temporal evolution is
\eq{
  P(\vec{x}, t+1) = \sum_{\vec{y}} M(\vec{x} | \vec{y}) P(\vec{y}, t),
}
or, in vectorial terms
\eq{
  \vec{P}(t+1) = \vec{M} \vec{P}(t).
}
The state of site $i$ at time $t+1$ depends on the state of its neighbourhood at time  $t$. The probability  that site $i$ takes the value $x_i$ at time $t+1$  given  its neighbourhood $(y_i)_{a_{ij}=1}$ at time $t$ is determined by the  (fixed) local transition probabilities $\tau(x_i| (y_i)_{a_{ij}=1})$ 
\eq{
  M(\vec{x}|\vec{y}) = \prod_i \tau(x_i| (y_j)_{a_{ij}=1}).
}

Again, we can define stochastic trajectories (or deterministic trajectories over a stochastic field) 
\meq{
 x_i(t+1) = [r_i(t) < & \tau(x_i| (y_j)_{a_{ij}=1})] = \\
 &\frac{1}{2} \left(1-\operatorname{sign}(r_i(t) -  \tau(x_i| (y_j)_{a_{ij}=1}))\right).
}
 
Deterministic cellular automaton can be considered as limit cases of PCA, where the transition probabilities $\tau$ are  either zero or one. 

\begin{table}[t]
\caption{\label{tab:DK} Transition probabilities of the Domany-Kinzel model. $S=s_{-1}+s_{+1}$, $X=x_{-1}+x_{+1}$ }
\begin{center}
\begin{tabular}{c|c|c|c|c|c}
\hline
\scriptsize $S$& \scriptsize $X$ & \scriptsize $\tau(1|S)$ &\scriptsize  $\tau(0|S)$&\scriptsize bond percolation&\scriptsize site percolation\\
\hline
-1 & 0 &  $w$ & $1-w$ &0 &0\\
0 & 1 & $p$ & $1-p$ & $p_b$ & $p_s$\\
1 & 2 & $q$ & $1-q$ & $p_b(2-p_b)$ & $p_s$\\
\hline
\end{tabular}
\end{center}
\end{table}

\subsection{Parallel Ising model}

For instance, we can define a parallel version  of the Ising model, for which 
\eq{
  M(\vec{s}'|\vec{s}) = \prod_i \tau(s'_i|(s_j)_{a_{ij} = 1}),
}
with $\tau$ given by Eq.~\eqref{isingprob}. 

In this case we can still have an asymptotic probability distribution if the interactions are symmetric (here they are for definition), but the asymptotic distribution is now~\cite{Derrida}
\eq{
  P^\text{eq} (\vec{s}) = \frac{1}{Z} \prod_i e^{\beta H s_i} \cosh\left(\sum_j \beta (H+J\sum_j  a_{ij} s_j)\right),
} 
where $Z$ is again the normalization constant. 

\begin{figure}[t]
\begin{center}
\begin{tabular}{|c|c|c|c|}
\hline
\includegraphics[height=0.1\textheight]{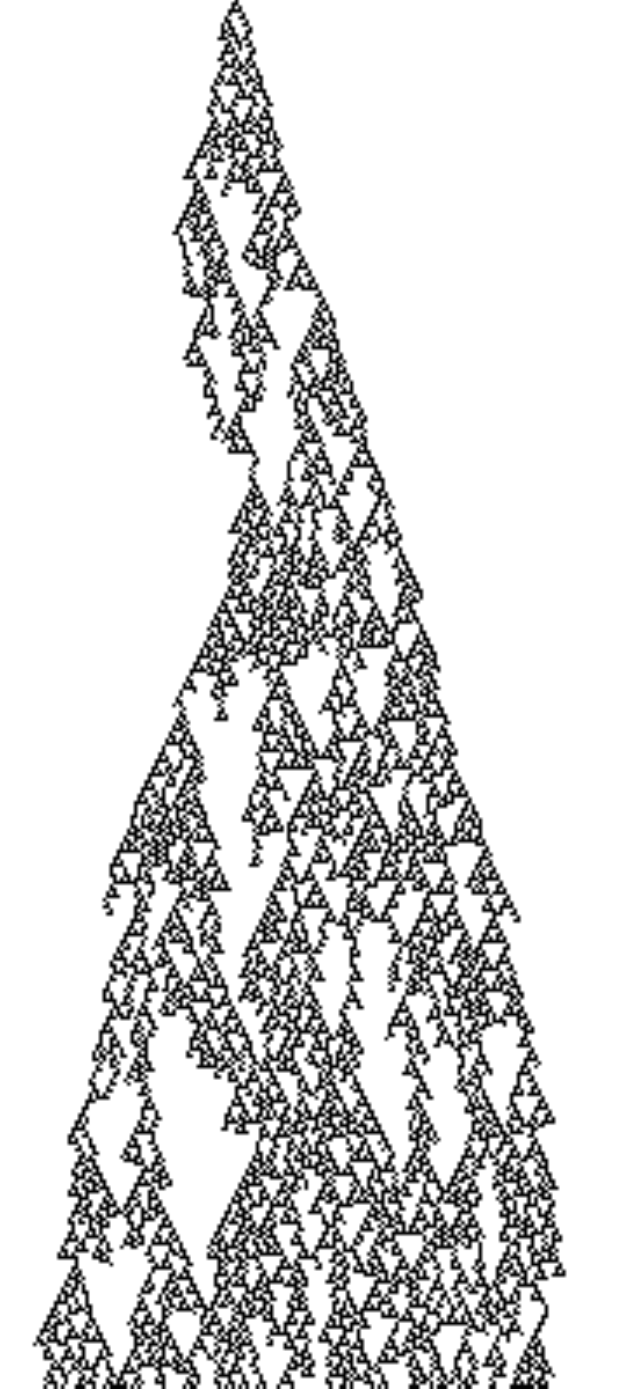} &
\includegraphics[height=0.1\textheight]{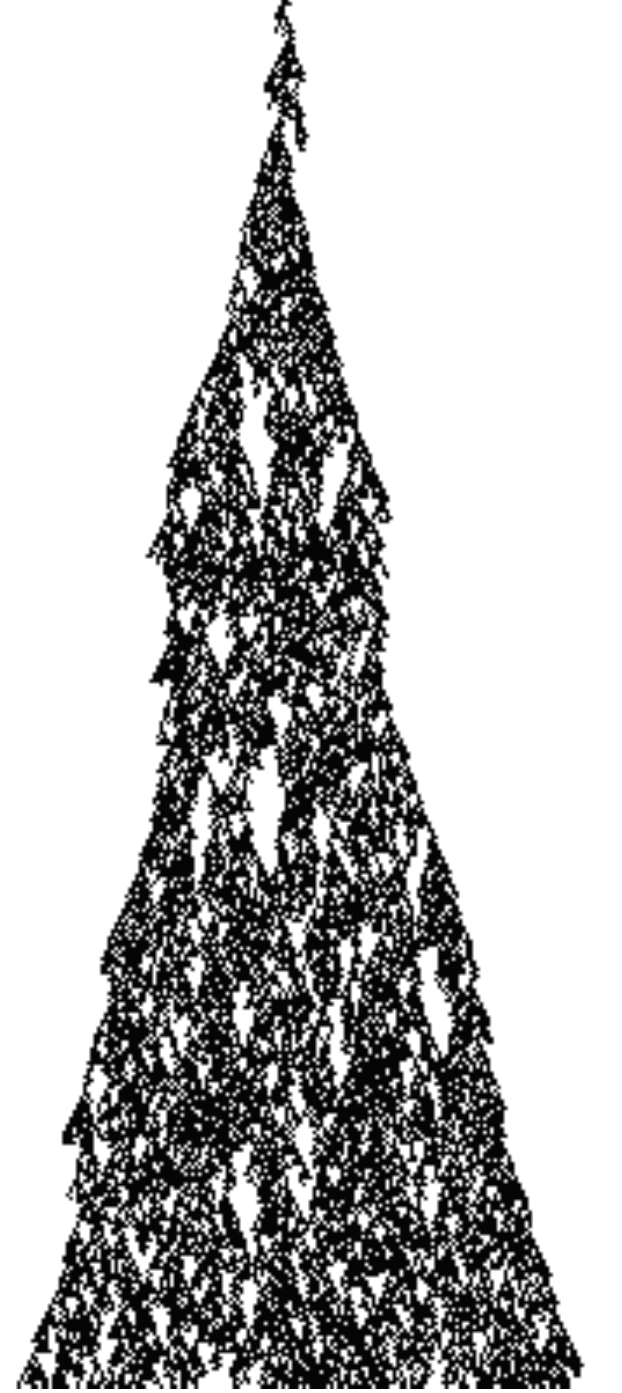} &
\includegraphics[height=0.1\textheight]{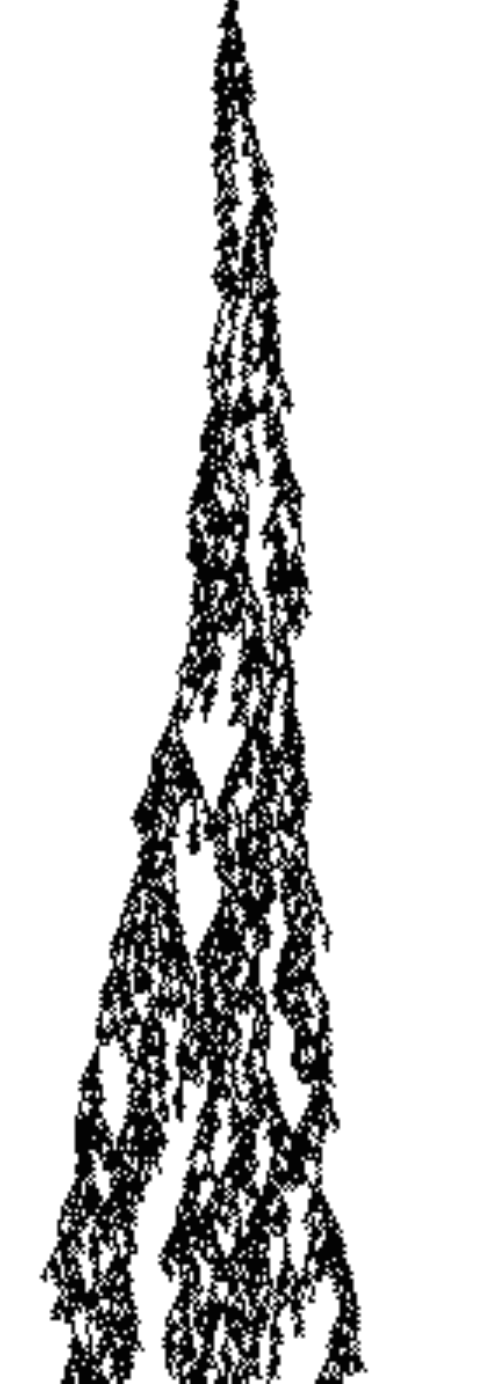} &
\includegraphics[height=0.1\textheight]{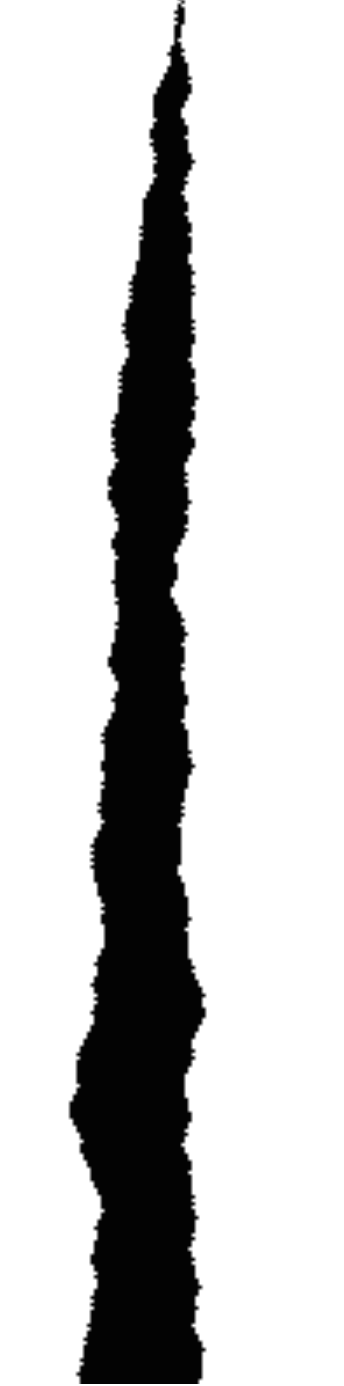} \\
\hline
\scriptsize
Dilution of DCA  90 &\scriptsize site percolation & \scriptsize bond percolation & \scriptsize Ising $T=0$\\
\scriptsize $p=0.81, q=0$ &\scriptsize $p=q=0.71$ &\scriptsize $p=0.64, q=0.87$ &\scriptsize $p=0.5, q=1$\\
\hline
\end{tabular}
\caption{\label{fig:DPpatterns} Typical patterns of the DK model. Space runs horizontally and time vertically, from top to bottom.}
\end{center}
\end{figure}


Notice that the transition probabilities of Eq.~\eqref{isingprob} do not depend on the previous value of the site $s_i$. If we apply them in parallel to all sites, at least in one dimension and with nearest-neighbour interactions, the lattice decouples in two noninteracting sublattices (for even $N$), so that $s'_i = f(s_{i-1}^t+s_{i+1} (t), r_i(t))$ It is an example of a totalistic PCA, that has been studied by Kinzel~\cite{Kinzel} and shows no phase transition.

\subsection{Domany-Kinzel model. Absorbing states.}

We can extend the parallel Ising example to a general case, on the same two-neighbours network, defining the three independent totalistic transition probabilities, as shown in Table~\ref{tab:DK}. This model has been studied by Domany and Kinzel~\cite{DK, Kinzel}, and can be considered the simplest model showing a phase transition. 

For generic values of $w$, $p$ and $q$ this model can be mapped onto an Ising model with a plaquette term~\cite{Kinzel} (we need another control parameter in addition to $H$ and $J$ since here we have three free probabilities), 
\eq{
  \mathcal{H}(\vec{S}) = -\sum_i s_i \left(H +  J (s_{i-1} +  s_{i+1}) + K  s_{i-1}s_{i+1} \right).
}

Denoting $h=\exp(-2H)$, $j = \exp(-4J)$, $k=\exp(-2K)$, we have $w=1/(1+hk/j)$, $p=1/(1+h/k)$, $q=1/(1+hjk)$ and therefore 
\meq{
  H &= \dfrac{1}{6} \log\dfrac{wpq}{(1-w)(1-p)(1-q)}, \\
  J &= \dfrac{1}{8} \log\dfrac{(1-w)q}{w(1-q)},\\
  K &= \dfrac{1}{6} \log\dfrac{w(1-p)q}{(1-w)p(1-q)}.
}

\begin{figure}[t]
\begin{center}
\includegraphics[width=0.53\columnwidth]{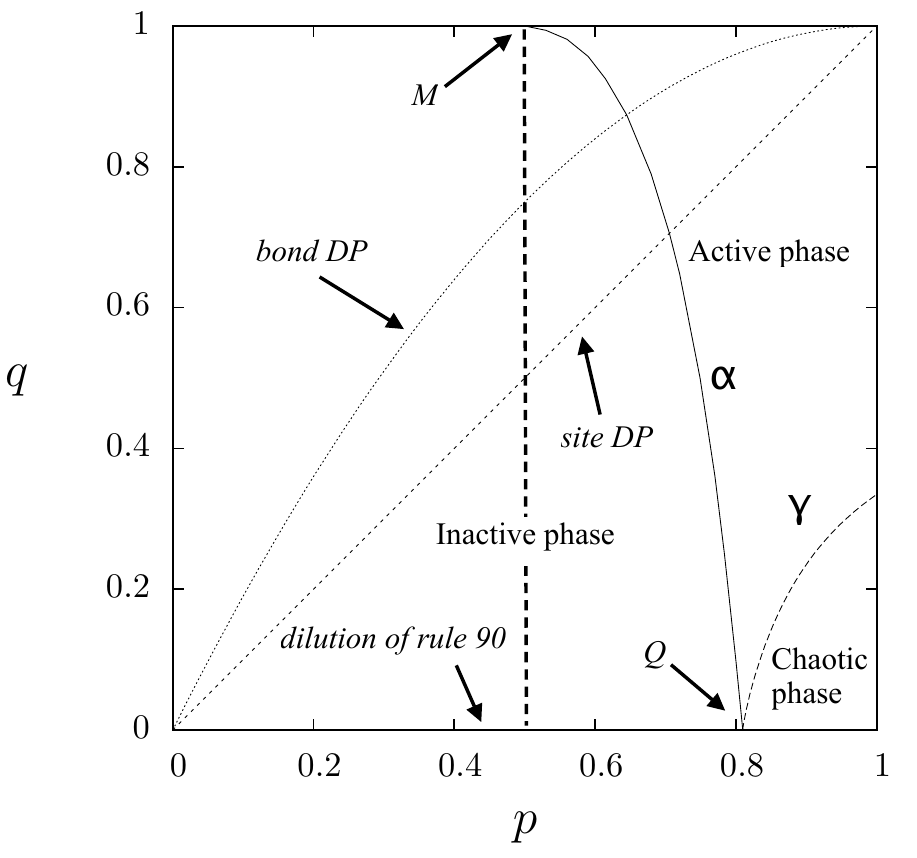}
\includegraphics[width=0.45\columnwidth, bb=0 -20 250 280]{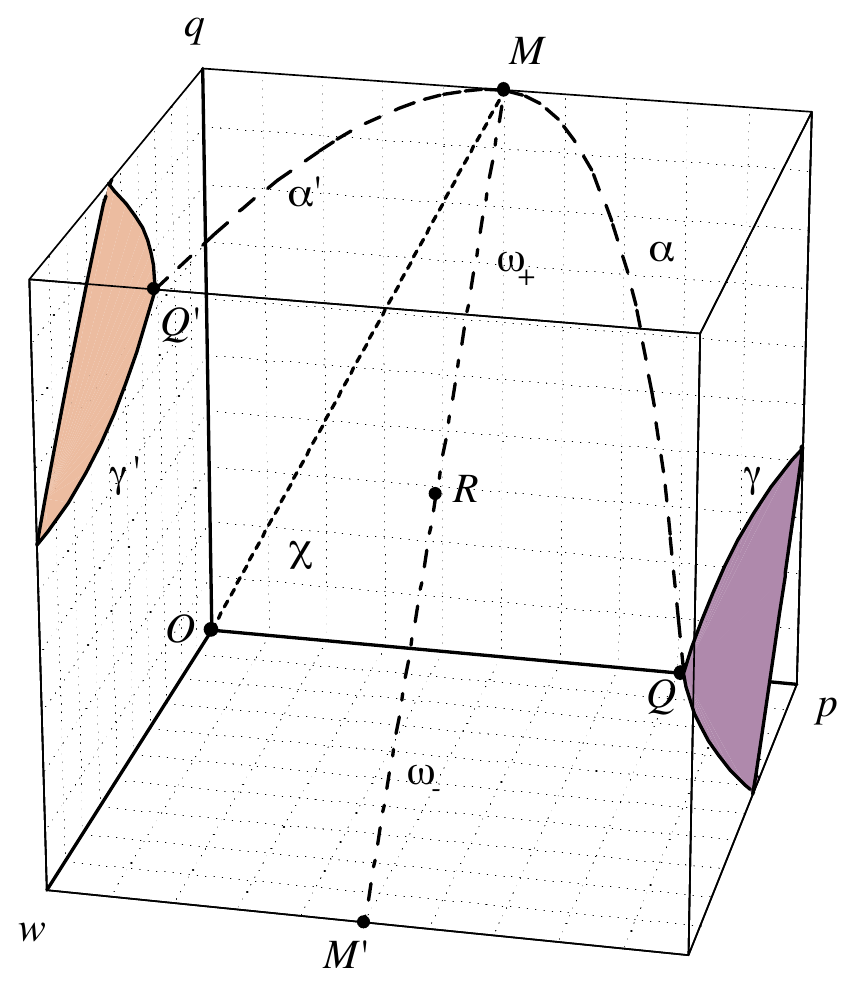}
\caption{\label{fig:DK} The phase diagram of the Domany-Kinkel model, $\alpha$ marks the density transition and $\gamma$ the damage transition. Left: the phase diagram for $w=0$. The dashed line marks the transition line for the simplest mean-field approximation. Right: the complete phase diagram. The  curves labelled $\alpha $ and $\alpha ^{\prime }
$ belong to planes $w=0$ and $w=1$ resp., and correspond to the density
phase transitions. The solid curves correspond to the intersection of the
damage critical surface (shaded) $\gamma $ and $\gamma ^{\prime }$ with the
boundaries of the cube. The dotted-dashed lines labelled $\omega _{+}$ and $%
\omega _{-}$ correspond to the existence line for the parallel Ising model for
positive and negative temperatures, resp. The points labelled $M$ and $%
M^{\prime }$ to the critical points of the parallel Ising model at zero temperature (compact DP), and
the point labelled $R$ to infinite temperature. The dotted line labelled $\chi 
$ corresponds to the damage in the parallel Ising model. }
\end{center}
\end{figure}
However, this model does not show any phase transition. 

If we set $w=0$ (by letting  the coupling take infinite values with suitable limits), we leave the equilibrium condition. In this limit the configuration $\vec{s} = -1$ becomes an absorbing state. We can also switch to the Boolean representation by setting $x_i = (s_i+1)/2$. In this representation the absorbing state is the configuration $\vec{x}=0$. It is called absorbing since it cannot be left by the dynamics once entered. The order parameter is here the ``density'' of ones 
\eq{
  c = \dfrac{1}{N} \sum_i x_i 
}

We can reformulate the phase transition in this new language: for finite $N$ there is always a probability $M(0|\vec{x})$ that brings any configuration to the absorbing state in one step. In the limit $N\rightarrow \infty$ and for a suitable value of the parameters $p$ and $q$ this probability goes to zero and the Markov matrix becomes reducible. It is composed by a submatrix $M_1$ that maps states ``near'' to 0 into 0 in  a few time steps, and a set of states with a non-vanishing density $c$

Again, one can speak of deterministic trajectories one that the stochastic field has been laid out. The evolution equation of the system is
\meq{
  x'_i =& [r^{(1)}_i(t) < p] (x_{i-1}(t) \oplus x_{i+1}(t)) \oplus \\
  &\qquad [r^{(2)}_i(t) < q] x_{i-1}(t) x_{i+1}(t)
}
where $\oplus$ is the XOR operation (sum modulus two). The two random numbers $r^{(1)}_i(t)$ and $r^{(2)}_i(t)$ may be the same or not, since the two conditions $(x_{i-1}(t) \oplus x_{i+1}(t)$ and $x_{i-1}(t) x_{i+1}(t)$ are never true at the same time (but this makes a difference for damage spreading, next Section).

In the language of trajectories, one can say that there are two attractors, the fixed point 0 and a ``chaotic'' attractor with $d>0$, each one with its own basin. More on absorbing phase transition in Ref.~\cite{nonequilibrium}.

\begin{figure}[t]
\centerline{\includegraphics[width=0.5\columnwidth]{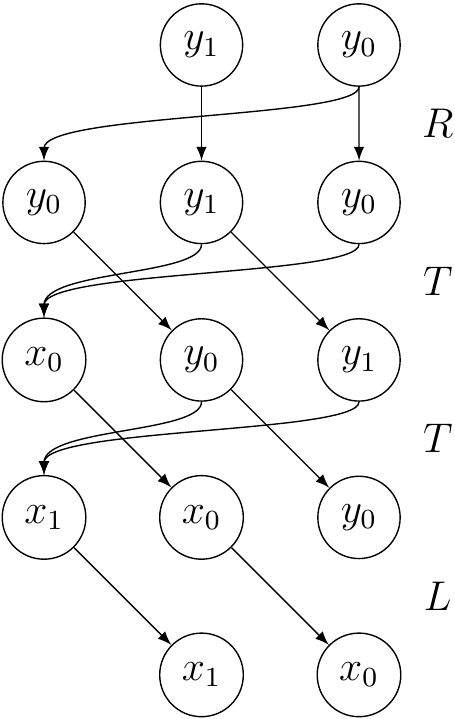}}
\caption{\label{fig:transfer} The application of the transfer matrix 
for the Domany-Kinzel cellular automata with $N=2$ and periodic boundary conditions.}
\end{figure}

For $w=q=0$ and $p=1$ we have the deterministic rule 90 in Wolfram's notation~\cite{Wolfram}, so the line $q=w=0$ corresponds to the dilution of rule 90. 

The DK model includes the Directed Percolation (DP) one~\cite{DP}, that can be formulated thinking to an infection process: an individual $i$ at time $t$ can get infected by its infected neighbours at the previous time step, with a probability  that depends on the number of infected neighbours (bond percolation) 
or not (site percolation), see Table~\ref{tab:DK}. Some typical patterns of the DK model starting from a single site are reported in Fig.~\ref{fig:DPpatterns}.

Let us develop the Markov approach for the DK model $N=2$ and periodic boundary conditions. 
The general equation 
\eq{
  P' (x_1, x_0) = \sum_{y_1,y_0} M(x_1,x_0|y_1, y_0)  P(y_1, y_0):  \; P' = M P,
}
is decoupled first in an expansion (for boundary conditions)
\eq{
  Q(y_2, y_1, y_0)  = P(y_1,y_0) [y_2=y_0]: \qquad Q = R P ,
 }
followed by two steps 
\meq{
    Q'(x_0, y_2, y_1)  &= \sum_{y_0} \tau(x_0| y_1, y_0) Q(y_2, y_1, y_0) : \; Q' = T Q, \\
    Q'(x_1, x_0, y_2)  &= \sum_{y_1} \tau(x_1| y_2, y_1) Q(x_0, y_2, y_1)  : \; Q' = T Q ,
}
and a contraction
\begin{figure}[t]
\begin{center}
\includegraphics[width=0.49\columnwidth]{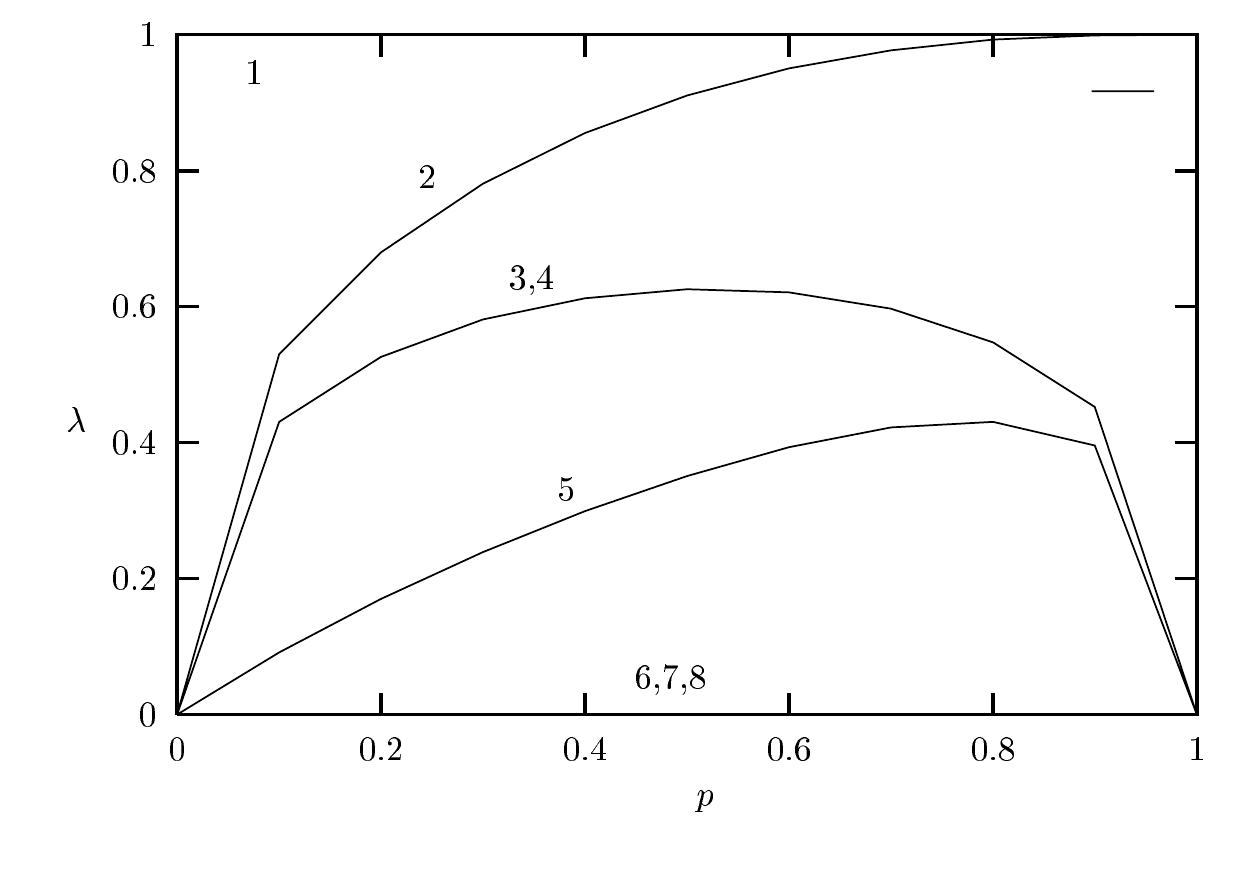}
\includegraphics[width=0.49\columnwidth]{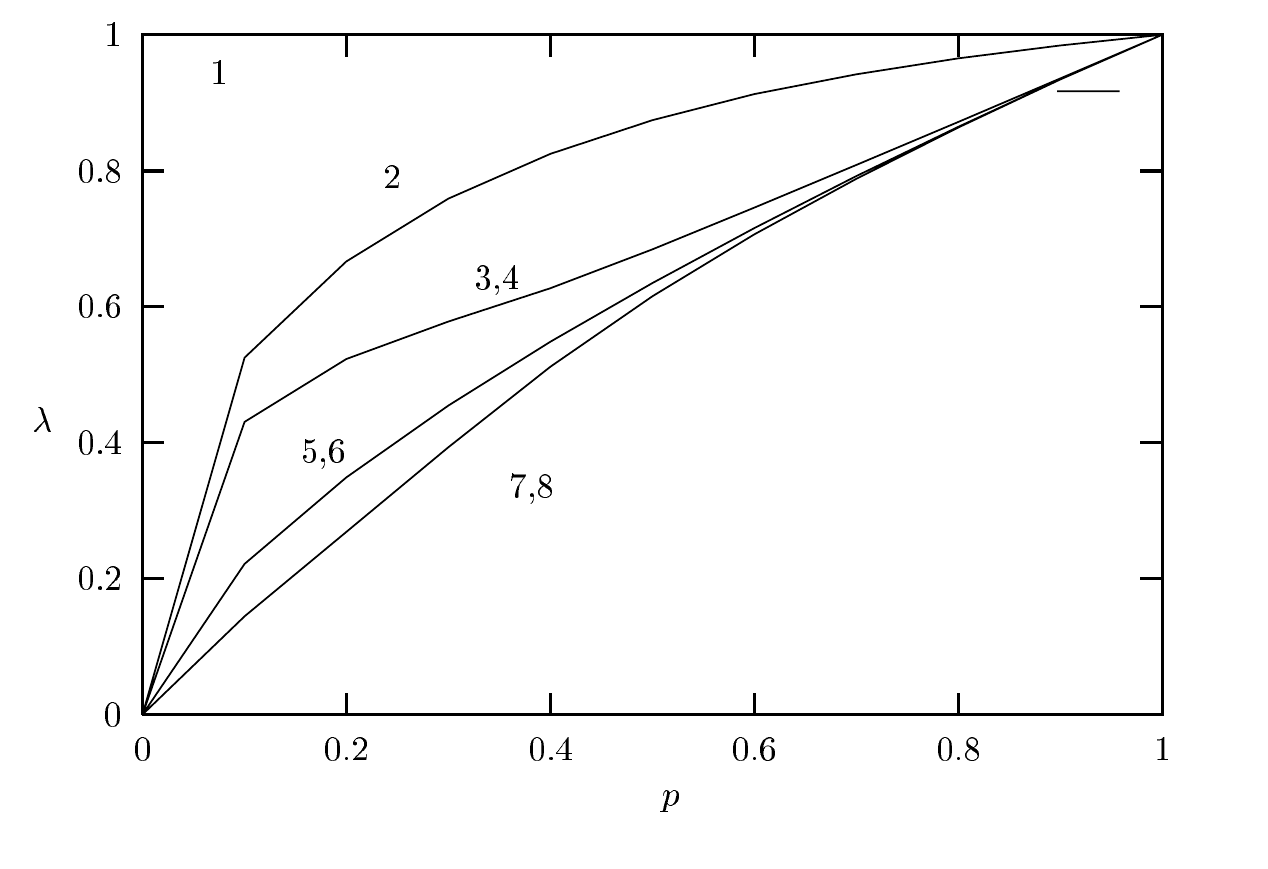}
\end{center}
\caption{\label{fig:degen}Modulus of the eigenvectors of the $N=2$ 
transfer matrix for $q=p$ (left) and $q=0$ (right).}
\end{figure}
\eq{
 P'(x_1, x_0) = \sum_{y_2}  Q'(x_1, x_0, y_2)  : \; P' = L Q ,
}
i.e.
\eq{
  M = L T^N R,
} 
as illustrated in Fig.~\ref{fig:transfer}, with 
{\scriptsize
\eq{
 T = \bordermatrix{
  ~& 000 & 001 & 010 & 011 & 100 & 101 & 110 & 111\cr
  000&  1 &  1 - p &  0 &  0 &  0 &  0 &  0 &  0 \cr
   001 &   0 &  0 &  1 - p &  1 - q &  0 &  0 &  0 &  0 \cr
   010 &   0 &  0 &  0 &  0 &  1 &  1 - p &  0 &  0 \cr
   011&   0 &  0 &  0 &  0 &  0 &  0 &  1 - p &  1 - q \cr
   100&    0 &  p &  0 &  0 &  0 &  0 &  0 &  0 \cr
   101 &   0 &  0 &  p &  q &  0 &  0 &  0 &  0 \cr
    110&  0 &  0 &  0 &  0 &  0 &  p &  0 &  0 \cr
    111&   0 &  0 &  0 &  0 &  0 &  0 &  p &  q
  }.
}
}

It is expected that, for $N$ large, the influence of the boundary conditions ($L$ and $R$ matrices) is not important, and elicoidal conditions can be used (i.e., pure $T$ matrices). 

The matrix $M$ and $T$ share the same eigenvectors, and the eigenvalues of $M$ are that of $T$ raised to the power $N$. 

At the transition, the second eigenvalue become degenerate with the first, as shown in Fig.~\ref{fig:degen}.

\subsection{Mean field approximation}
\label{sec:meanfield}

In order not to use a heavy notation, let us obtain this approximation using the DK model, assuming that a site  $i$ at time $t+1$ is connected to sites $i$ and $i+1$ at time $t$ (i.e., using the skewed lattice of Fig.~\ref{fig:neigh}). 

The evolution equation for the probability distribution is 
\meq[DK]{
  &P(x_1,x_2, \dots, x_N; t+1) =\\
  & \sum_{y_1,y_2, \dots, y_N} \left(\prod_i \tau(x_i|y_i, y_{i+1}\right) P(y1,y_2, \dots, y_N; t),
}
considering appropriate boundary conditions (e.g., periodic). We can obtain the reduced probabilities $\pi_\ell(x_i, \dots, x_\ell;t)$ by summing $P(x_1,x_2, \dots, x_N; t+1)$ over all $i>\ell$. If the system is translation invariant, one obtains the same result summing an any consecutive set of variables. Since $\sum_{x_i} \tau(x_i|y_i, y_{i+1}=1$ for all $x_i$, we can then sum over $y_{i+2}, \dots, y_N$, obtaining
\meq{
\pi_1(x_1, t+1) &= \tau(x_1|y_1,y_2) \pi_2(y_1, y_2;t),\\ 
\pi_2(x_1,x_2, t+1) &=\tau(x_1|y_1,y_2)\tau(x_2|y_2,y_3) \pi_3(y_1, y_2y_2;t),\\ 
\dots
}
i.e., a hierarchy of equations that are equivalent to Eq.~\eqref{DK}.

If the correlation length $\xi$ is
less than $N$, two cell separated by a distance greater that $\xi$ are
practically independent. The system acts like a collection of subsystems each
of length $\xi$ (this is why ergodicy and selfaveraging holds far from the transition). Since $\xi$ is not known a priori, one assumes
a certain correlation length $\ell$ and computes the quantity of
interest. By comparing the values of these quantities
with increasing $\ell$ generally a clear scaling law appears, 
allowing to extrapolate the results to the case $\ell\rightarrow\infty$.

The very first step is to assume $\ell=1$. In this case we can simply factorize $\pi_2(x_1,x_2)=\pi_1(x_1)\pi_1(x_2)$. By calling $c=\pi_1(1;t)$ ($1-c=\pi_1(0;t)$), $c'=\pi_1(1;t+1)$ and using the transition probabilities of Table~\ref{tab:DK} with $w=0$, one gets
\eq{
  c' = 2pc(1-c) + q c^2.
}
The fixed points of this map ($c'=c$) are $c=0$ and $c=2p/(2p-q)$. There is a change of stability from $c=0$ (the absorbing state) to $c>0$ for  $p_c=1/2$. 
As shown in Fig.~\ref{fig:DK}-left, this approximation is quite rough. 

In the mean-field approximation, we have for the bond percolation the line $q=p(2-p)$, and for the site percolation the line $q=p$, as shown in Fig.~\ref{fig:DK}-left.

There are two ways of extending the above approximation. The first is still to
factorize the cluster probabilities at single site level but to consider more
time steps, for instance obtaining $\pi_1(t+2)$ in terms of $\pi_3(t)$ and then factorizing $\pi_3$ in terms of $\pi_2$. 
The map
is still expressed as a polynomial of the density $c$. The
advantage of this method is that we still work with a scalar (the density),
but in the vicinity of a phase transition the convergence towards the
thermodynamic limit is very slow. 

\begin{figure}[t]
\begin{center}
\includegraphics[width=0.75\columnwidth, bb=20 20 320 250]{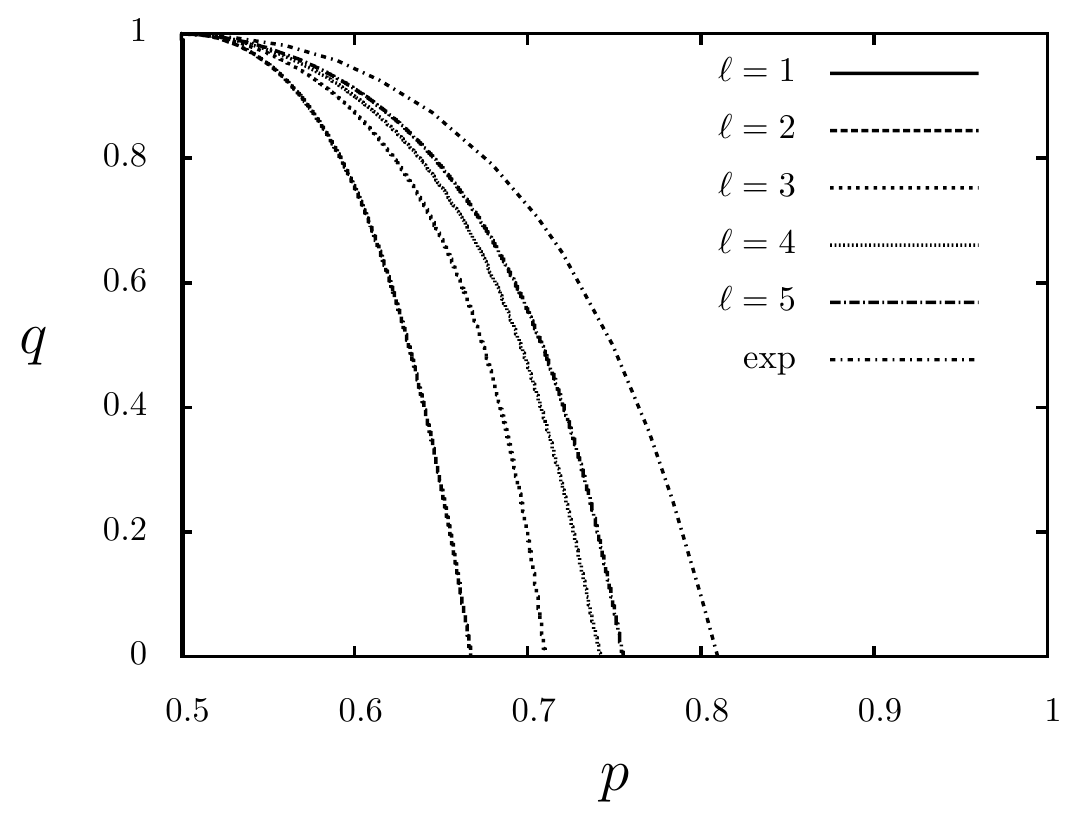}
\caption{\label{fig:localapprox} Local structure approximation for the DK model, with several values of length $\ell$. The case $\ell=1$ is the simplest mean-field approximation and corresponds to the line $p=1/2$. The line marked 'exp' corresponds to numerical simulations as in Fig.~\ref{fig:DK}.}
\end{center}
\end{figure}

The second approach, sometimes called \textit{local structure
approximation}~\cite{localstructure}, is a bit more complex. Let us start from
the generic $\ell$ cluster probabilities $\pi_\ell$. We generate the $\ell-1$ cluster
probabilities $\pi_{\ell-1}$ from $\pi_\ell$ by summing over one variable,
\eq{ 
	\pi_{\ell-1}(x_1, \dots, x_{\ell-1}) = \sum_{x_\ell} \pi_{\ell}(x_1,
	\dots,  x_{\ell-1}, x_\ell). 
} 
The $\ell+1$ cluster probabilities are
generated by using a Bayesian estimation
\eq{ 
	\pi_{\ell+1}(x_1,x_2,\dots,  x_\ell, x_{\ell+1})=
		\dfrac{\pi_{\ell}(x_1, \dots, x_\ell) \pi_{\ell}(x_2, \dots,
		x_{\ell+1})}{ \pi_{\ell-1}(x_2, \dots, x_{\ell})}. 
} 
Finally, one is back
to the $\ell$ cluster probabilities by applying the  transition probabilities
\eq{
	\pi'(x_1, \dots, x_\ell)= \sum_{y_1,\dots,y_{\ell+1}}
	\prod_{i=1}^l\tau(x_i|y_i, y_{i+1}). 
}
This last approach has the disadvantage that the map lives in a
high-dimensional ($2^\ell$) space, but the results converges much better in
the whole phase diagram. 

This mean field technique can be considered an application of the
transfer matrix concept to the calculation of the the eigenvector (asymptotic probability distribution) corresponding
to the maximum eigenvalue (fundamental or ground state), by means of the iteration of the matrix.

\subsection{Asynchronism of DCA}

An unexpected phase transition occurs with an increasing level of asynchronism of some DCA rule~\cite{Fates}. 
 Let us denote by $f(x_{i-1}, x_i, x_{i+1})$ the deterministic rule. The evolution equation of its dilution  is 
\eq{
  x'_i = x_i \oplus [r_i(t) <  (1-p)] \bigl(x_i \oplus f(x_{i-1}, x_i, x_{i+1})\bigr). 
}
With probability $1-p$ the site follows the rule $f$, and with probability $p$ it keeps its old value. 

\begin{figure}[t]
\begin{center}
\includegraphics[width=0.9\columnwidth]{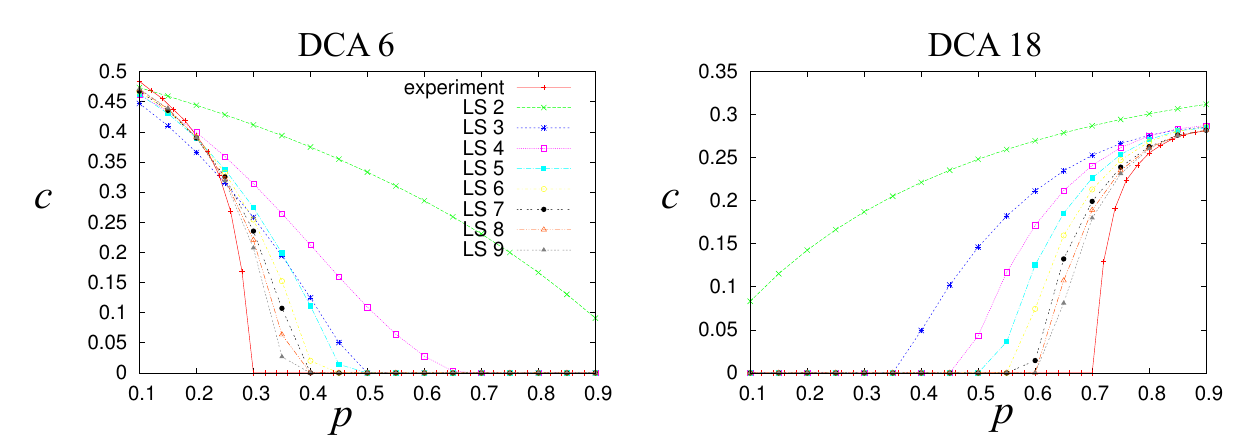}
\caption{\label{fig:async} DCA dilution phase transition for two Elementary Cellular Automaton rule 6 and 18 in Wolfram notation~\cite{Wolfram}, comparisons between numerical simulations and the local structure approximation (from Ref.~\cite{FuksFates}). In the $y$ axis the asynchronism parameter $p$.}
\end{center}
\end{figure}

Examples of phase transitions are shown in Fig~\ref{fig:async}. 

An unexpected fact is that the simplest mean-field approximation completely fails for this problem. Indeed, we have 
\eq{
  c'  = p c + (1-p) \sum_{a,b,c=0}^1 f(a,b,c) c^{a+b+c}(1-c)^{3-a-b-c}
}
and for the stationary state $c'=c$ one gets
\eq{
 c = \sum_{a,b,c=0}^1 f(a,b,c) c^{a+b+c}(1-c)^{3-a-b-c}
}
i.e., the mean-field approximation of the deterministic rule, without any dependence on $p$. Increasing the order of the mean-field approximation (local structure approximation), one can approximate the actual phase transition behaviour~\cite{FuksFates}, as shown in Fig.~\ref{fig:async}.

\subsection{Damage spreading}

We have said that the large-time distribution $\vec{x}(T)$ depends in general on the random field and the initial conditions $\vec{x}(0)$, although, for large $N$, the observables like the density does not depend on them due to ergodicity and self-averaging. Actually, we can check the dependence on the initial conditions by considering the evolution of an initial difference between two replicas, evolving on the same random field, and looking at the difference (or damage) $z_i = x_i \oplus y_i$, 
\meq{
x'_i &=  [r^{(1)}_i(t) < p] \bigl(x_{i-1}(t) \oplus x_{i+1}(t)\bigr) \oplus \\
 &\qquad  [r^{(2)}_i(t) < q] x_{i-1}(t) x_{i+1}(t), \\
y'_i &= [r^{(1)}_i(t) < p] \bigl(y_{i-1}(t) \oplus y_{i+1}(t)\bigr) \oplus\\
&\qquad   [r^{(2)}_i(t) < q] y_{i-1}(t) y_{i+1}(t), \\
z'_i &= x'_i \oplus y'_i = [r^{(1)}_i(t) < p]\bigl(z_{i-1}(t) \oplus z_{i+1}(t)\bigr)\oplus\\
&\qquad   [r^{(2)}_i(t) < q] \cdot \\
 & \bigl((z_{i-1}(t) z_{i+1}(t) \oplus z_{i-1}(t) x_{i+1}(t) \oplus x_{i-1}(t) z_{i+1}(t) \oplus\\
 &\qquad  x_{i-1}(t) x_{i+1}(t)\bigr).
}
Since now the two conditions can occur at the same time, there is a difference in the evolution if one uses one or two random numbers per site (or if they are otherwise correlated). Looking only at the evolution of the difference  $\vec{z}$, the evolution of the $x$ replica (which is not affected by $z$) is just another  field (although it is not fully random). The quantity $z$ shows another phase transition (Fig.~\ref{fig:DK}) that characterizes the dependence on the initial condition: in one phase the difference goes to zero, meaning that all initial conditions will follow after a transient time the same trajectory, only depending on the stochastic field. In the other phase, the system maintains forever some memory of the initial condition. 

This phase transition also belongs to the directed percolation universality class. It is possible to approximately map the density phase transition onto the damage one~\cite{Bagnoli-damage}.

\subsection{Grassberger's conjecture}

We may have different scenarios, according to the degree of unpredictability of the system. Chaotic systems are expected to amplify the distance between replicas. For a value of $p$ slightly below the synchronization threshold, some patches may synchronize for some time, after which they will separate. This picture resembles that of a growing inter-face that may stay pinned to local traps.

From field theory studies,such a behaviour is denoted multiplicative noise (MN) and is equivalent to the behavior of the ``bounded'' Kardar-Parisi-Zhang equation, which describes the behaviour of a growing surface that tends to pin and is pushed from below~\cite{KardarZhang,MunozHwa,TuGrinsteinMunoz} .  On the other hand, stable systems have a negative MLE. So, replicas should naturally synchronize once their distance is (locally) below the threshold of validity of linear analysis. However, when the local difference is large, non-linear terms may maintain or amplify this distance. In this case synchronized patches may be destabilized only at the boundaries. Again, theoretical studies associate such a behaviour to that of directed percolation (DP)~\cite{DP}.

Grassberger~\cite{Grassberger} conjectured that every system with 
\begin{itemize}
\item Local interactions (not long-range nor random)
\item Asymmetric, stable absorbing state (not unstable for infinitesimal perturbations, nor with many absorbing states with the same probability)
\end{itemize}
does belong to the Directed percolation universality class.

\subsection{A richer phase diagram: the BBR model}

\begin{figure}[t]
\begin{center}
\includegraphics[width=0.48\columnwidth]{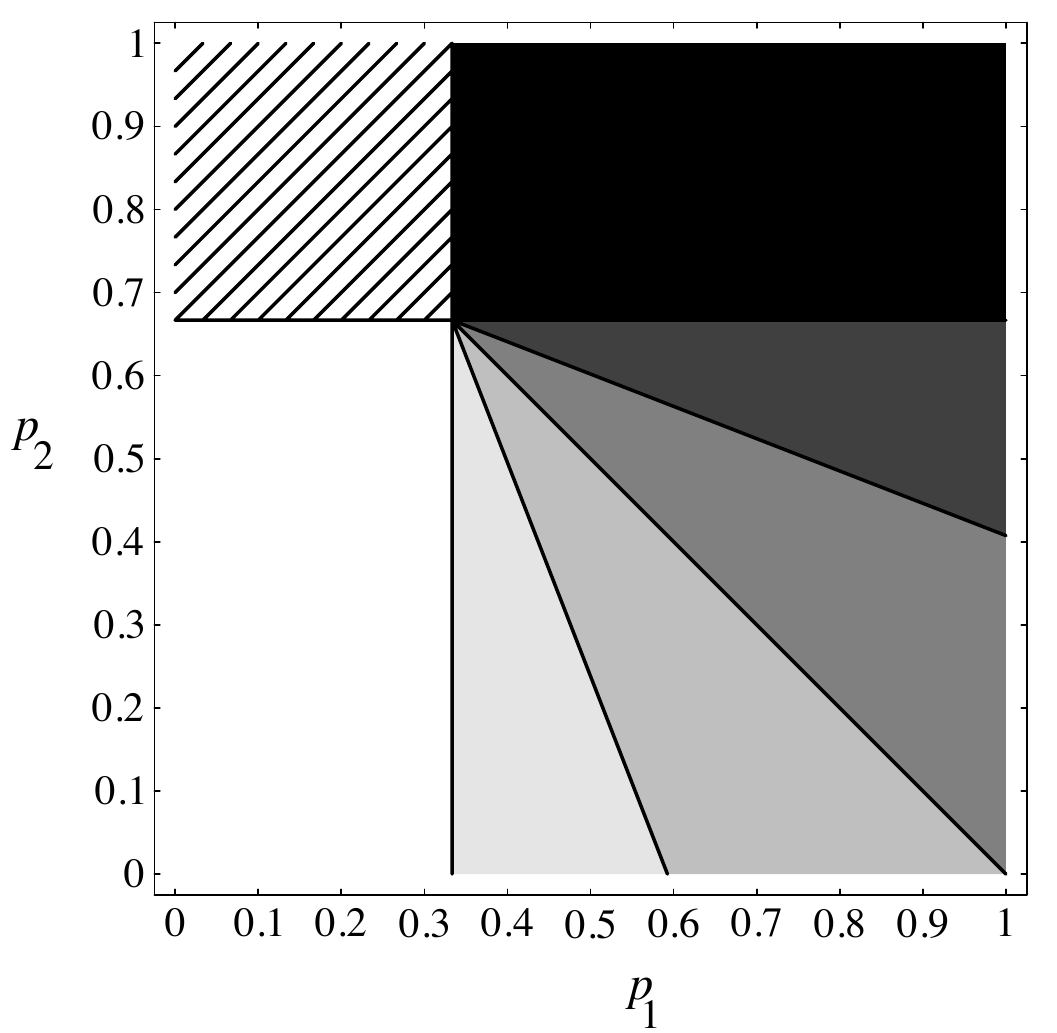} 
\includegraphics[width=0.48\columnwidth]{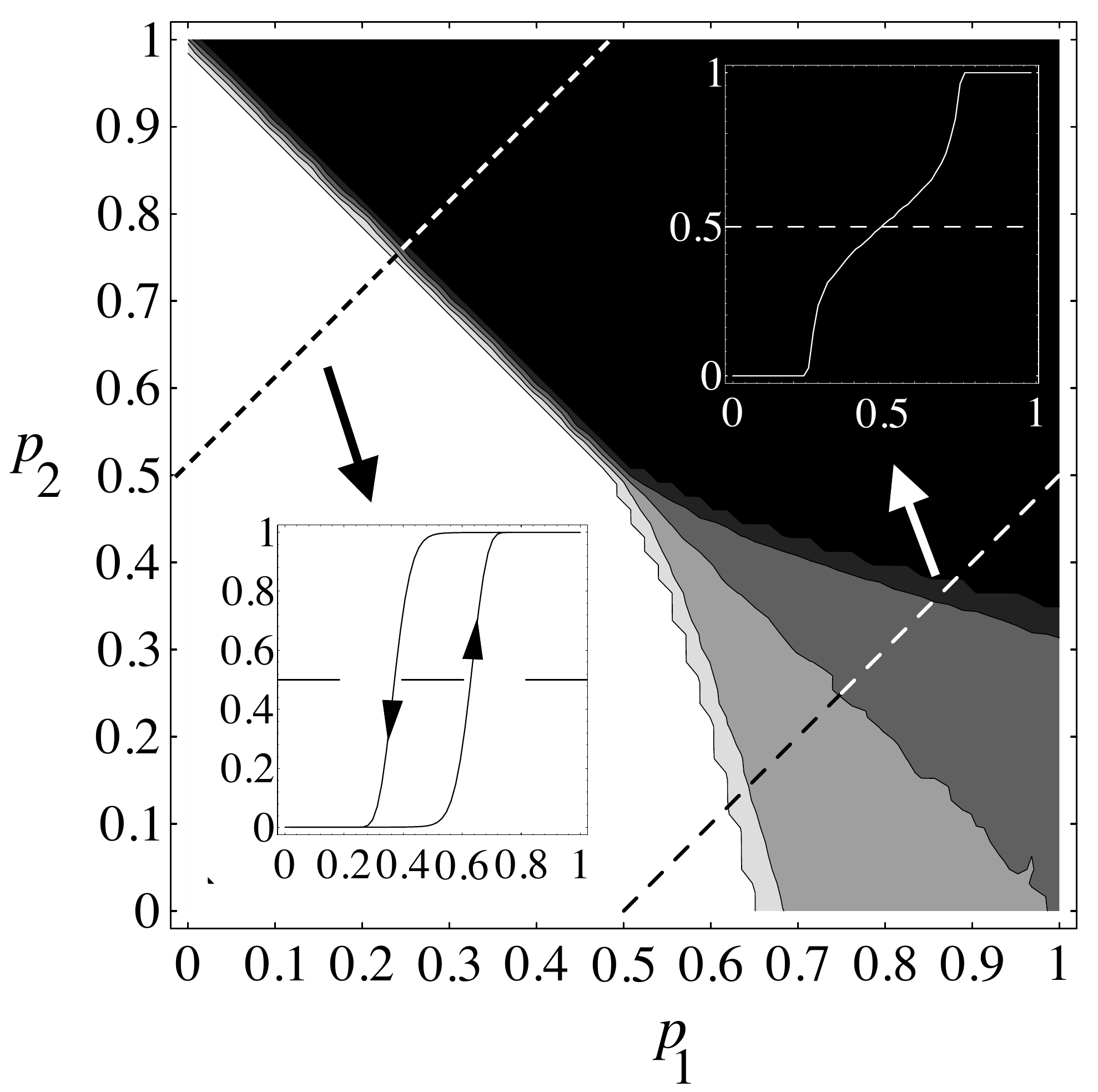} 
\caption{\label{fig:BBR} Phase transition diagrams of the BBR model (colour code: white=0, black=1). Left: mean field phase diagram for the density.  Right: numerical phase diagram of the density, in the inset the variation of the density when cutting the phase diagram; the hysteresis inset at bottom right is obtained by setting $w=10^{-4}$, $T=500$.  Numerical simulations with $N=T=10^4$. }
\end{center}
\end{figure}

The DK model is quite useful for studying nonequilibrium phase transitions due to its simplicity. In order to explore other types of transitions beyond DP, let us introduce the BBR model~\cite{BBR}, that is a 3-input cellular automata with two absorbing states. It is a totalistic automaton, meaning the the transition probability depends on the sum $S$ of the states in the neighborhood, with $0\le S\le 3$. The BBR transition probabilities $\tau(x'|S)$ are $\tau(1|0)=w$, $\tau(1|1)=p_1$, $\tau(1|2)=p_1$, $\tau(1|3)=1-w$
By setting $w=0$, the states $0$ and $1$ are absorbing, and on the line $p_1=1-p_2$ the system is symmetric for the inversion $1\leftrightarrow 0$. 

As can be seen in Fig.~\ref{fig:BBR}, we have here for high-$p_1$ and low-$p_2$ two DP transitions reminiscent of the DK model. The two lines met at about $p_1=p_4=0.5$ ($p_1=1-p_2=1/3$ in the mean-field approximation). In this point the universality class changes to that of parity conservation, as predicted by the Grassberger hypothesis (here the basins of the two attractors are symmetric. in the  low-$p_1$, high-$p_2$ part of the diagram, we have a first-order transition: the two absorbing states are stable (as predicted by the mean-field analysis) and we can investigate the nature of an hysteresis cycle. In order to do that, we have to have a little  the states 0 and 1 


%
%
%
%

\begin{figure}[t]
 \begin{center}
   \includegraphics[width=0.55\columnwidth]{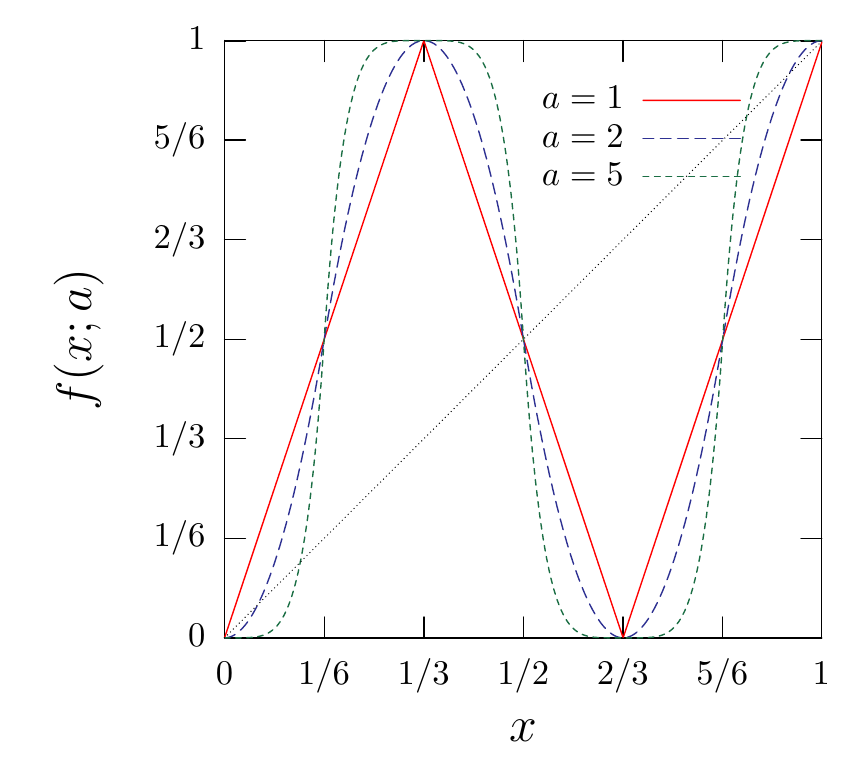}
  \includegraphics[width=0.38\columnwidth, bb=-10 -50 260 280]{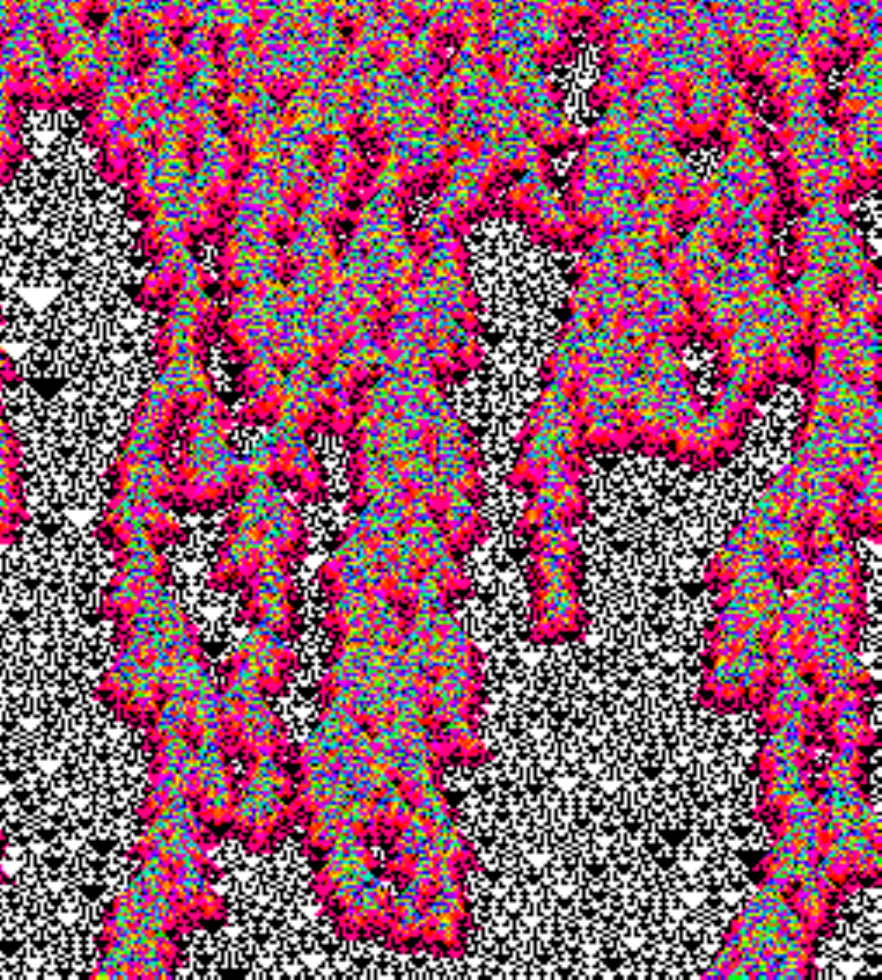}
 \caption{\label{fig:f} (left) The graph of $f(x;a)$ for three values of $a$. (right) Space time pattern of the CML of Eq.~\eqref{eq:f} with $a=1.9$ and $N=256$ drawn horizontally for a total time of $T=300$ time steps drawn vertically from top to bottom. The initial configuration $x(0)$ is chosen randomly. The color code assigns white (black) whenever $x_i(t)=0 (1)$ and a rainbow color scale for other values of $x_i(t)$ starting with red for values near zero. Patches of CA behavior (rule 150) appear after a short transient and will eventually fill the whole pattern.}
 \end{center}
 \end{figure}

\subsection{Synchronization}

An example of phase transition is given by the replica synchronization problem. The idea is the following: take two replicas of a system, either driven by a deterministic or a stochastic dynamics. Let one system evolve by itself, and ``push'' the other towards the first. If the pushing is enough, the system will synchronize. A simple illustration is the following. Take a continuous map $  x' = f(x)$,
and construct the synchronization mechanism
\meq{
  x' &= f(x),\\
  y' &= (1-p) f(y) + p f(x),
}
for $p=0$ the two systems are completely disconnected, and if the map $f$ is chaotic, they stay well separated. For $p=1$ the two system are identically the same. There is a critical value $p_c$ such that the distance $\delta = |x-y|$ goes to zero. For small distance, $\delta$ evolves as
\eq{
  \delta' = (1-p) |f(y) -f(x)| \simeq (1-p) \left|\dfrac{\mathrm{d} f(x)}{\mathrm{d}x}\right| \delta
}
and thus 
\meq{
  \delta(t) &= (1-p)^t \delta(0)\prod_{t'} \left|\dfrac{\mathrm{d} f(x(t'))}{\mathrm{d}x}\right| \\
  &= (1-p)^t \delta(0) \exp\left(\sum_{t'}\log \left|\dfrac{\mathrm{d} f(x(t'))}{\mathrm{d}x}\right|\right) \\
  &= \delta_0 \exp((\log(1-p) + \lambda) t),
}
where $\lambda$ is the Lyapunov exponent of the map. Thus, when $\delta(t) =\delta(0)$ (the synchronization threshold), $p_c  = 1-\exp(-\lambda)$, and this relates the synchronization threshold to the chaotic properties of the map. 

This mechanism can be extended in several ways to extended systems (coupled map lattices and cellular automata). For reference, consider the following generic coupled system
\eq{
   x'_i = f(g(x_{i-1}, _i, x_{i+1})),
}
where $g$ defines the coupling.
One can use a homogeneous ``pushing'', i.e., use the same $p$ for all sites, or, at the other extreme, a all-or-none pushing, i.e., choose a fraction $p$ of sites to be completely synchronized and leave the other unperturbed. 

Using the first mechanism, one again relates the synchronization threshold to the maximum Lyapunov exponent of the system. However, it is questionable if this exponent really captures the chaotic properties of an extended system. For instance, let take $f$ chaotic and $g(a,b,c) = \varepsilon(a+c) + (1-\varepsilon) b$, i.e., a diffusive coupling. The Lyapunov exponent $\lambda (\varepsilon)$ in general decreases with $\varepsilon$, since the coupling acts like a constraints (a kind of surface tension). Thus $lambda$ is maximum for $\varepsilon=0$, but in this case the chaos does not spread on the lattice. 

On the contrary, the all-or-none (``pinching'') synchronization mechanism shows that the case in which synchronization is most difficult is for $\varepsilon\simeq 1/3$, which is what one intuitively  expects. Moreover, we can apply this synchronization mechanism also to cellular automata, providing that the two replicas evolve using the same random numbers (field). It is possible to show that in this case one can develop a concept of Boolean derivative for such a discrete systems, and obtain  an equivalent of the maximum Lyapunov exponent, which is related to the pinching synchronization threshold~\cite{BooleanDerivatives,Bagnoli-lyapca}. 

The synchronized state is an example of absorbing state, but clearly in real cases one rarely expect to find a complete synchronization: the evolution may be influenced by noise, or the two replicas can be slightly different.

We can test this hypothesis using  the map
\eq[eq:f]{
 f(x;a)=
 \begin{cases}
  (6x)^a/2		& 0\leq x < 1/6,\\
  1-|6(1/3-x)|^a/2	& 1/3\leq x < 1/2,\\
  |6(x-2/3)|^a/2	& 1/2\leq x < 5/6,\\
  1-(6(1-x))^a/2	& 5/6\leq x < 1,
 \end{cases}        
}
where $1\le a<\infty$ (see Fig.~\ref{fig:f}-left), see Ref.~\cite{Bagnoli-smoothCML}. 
This map that has the advantage of reducing to the DCA rule 150 for $a$ large, and to a chaotic map from $a$ small. For $a\gtrsim 1.81$ (stable chaos) one observes a transient chaos, with positive Lyapunov exponent, followed by a cellular automata pattern. One may wonder about the unpredictability of such map: in the chaotic phase an infinitesimal damage will amply, while in stable chaos phase infinitesimal damages are absorbed (and thus the word ``stable'') but finite ones spread (and thus the word ``chaos''). The synchronization procedure applied to a lattice of such maps indeed shows that a certain effort is needed even in the ``stable'' phase to get the synchronization. In agreement with  the Grassberger conjecture, one finds the the synchronization phase transition for $a<1$ do belongs to the MN universality class, while for $a\gtrsim 1.81$ 

Such a behaviour is not limited to systems that reduce to DCA, see Ref.~\cite{Bagnoli-synchrononchaos} for an example.

\begin{figure}[t]
 \begin{center}
  \includegraphics[width=0.48\columnwidth]{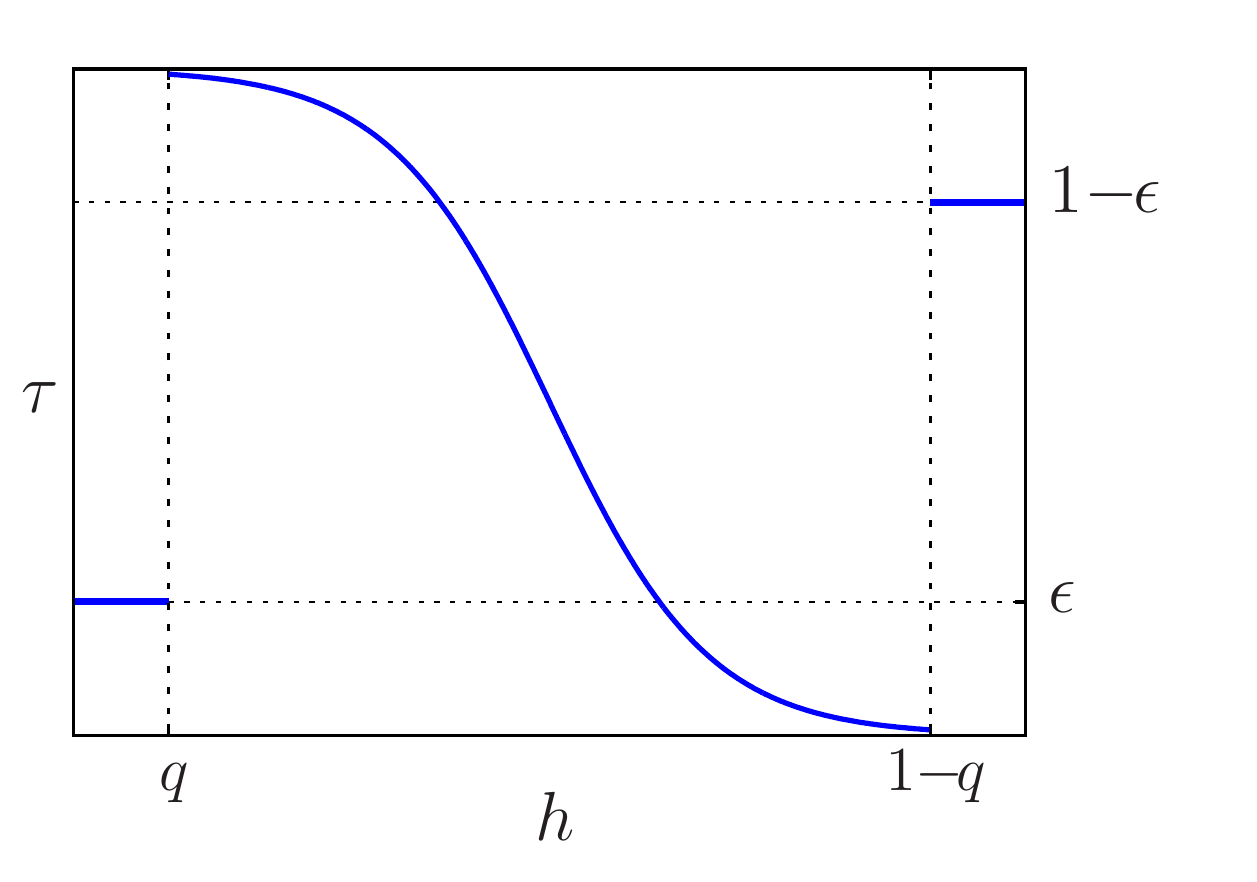}
   \includegraphics[width=0.48\columnwidth]{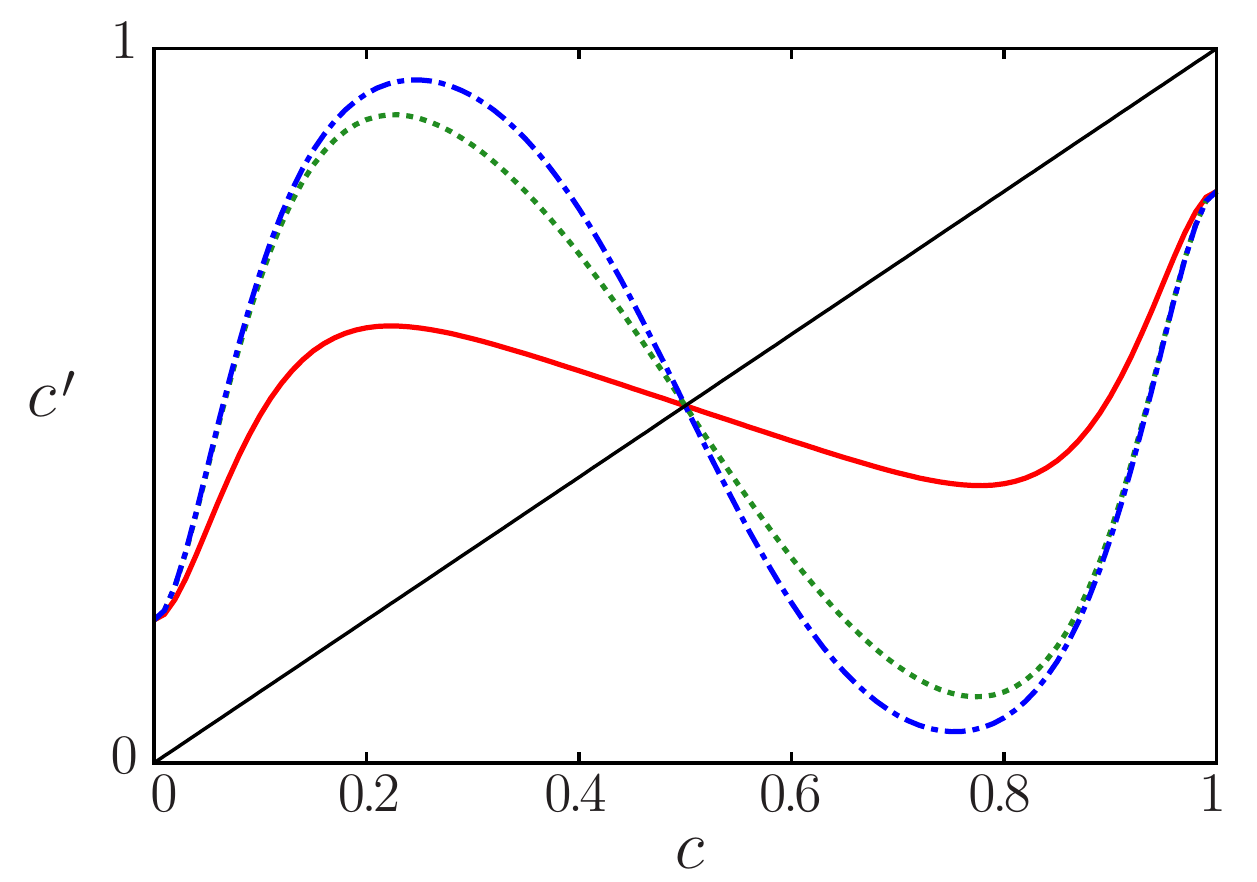} 
 \end{center}
  \caption{\label{fig:mfeq} (left)  The transition
    probability $\tau(h)$ given by Eq.~\eqref{tau} with $J=-3$,
    $k=20$, $q=0.1$, and $\varepsilon = 0.2$. (right) Graphs of the mean
    field map, Eq.~\eqref{eq:mf} for different values of $J$ and
    $k=20$. From bottom to top for
    $c<1/2$, $J=-0.5$ (red, lower line), $J=-3.0$ (green, middle line), and $J=-6.0$
    (blue, upper line).} 
\end{figure}

\subsection{Topology and chaotic phase transitions}

Up to now we have not investigated the influence of the topology, i.e., of the connections defined by the adjacency matrix $a_{ij}$. It is well known that if we replace a regular lattice with a random network of the same connectivity, the global behaviour becomes that of the mean-field, since in this way correlations are disrupted. 

We can study the influence of the topology by adopting the Watts-Strogatz rewiring mechanism~\cite{smallworld}: start with a regular lattice of connectivity $k$ in 1D and, for each site, rewire at random a fraction $ p$ of incoming links.

In order to show the effects of the mechanism and also to present a new type of phase transition, let us consider a cellular automaton whose mean-field approximation is chaotic. This model has been developed originally as an opinion formation mode~\cite{Bagnoli-opinion}

The average local opinion or social pressure $h_i$, is defined by
\eq{
 h_i =\dfrac{\sum_{j} a_{ij}s_j}{k}.
}

\begin{figure}[t]
 \begin{center}
   \includegraphics[width=0.48\columnwidth]{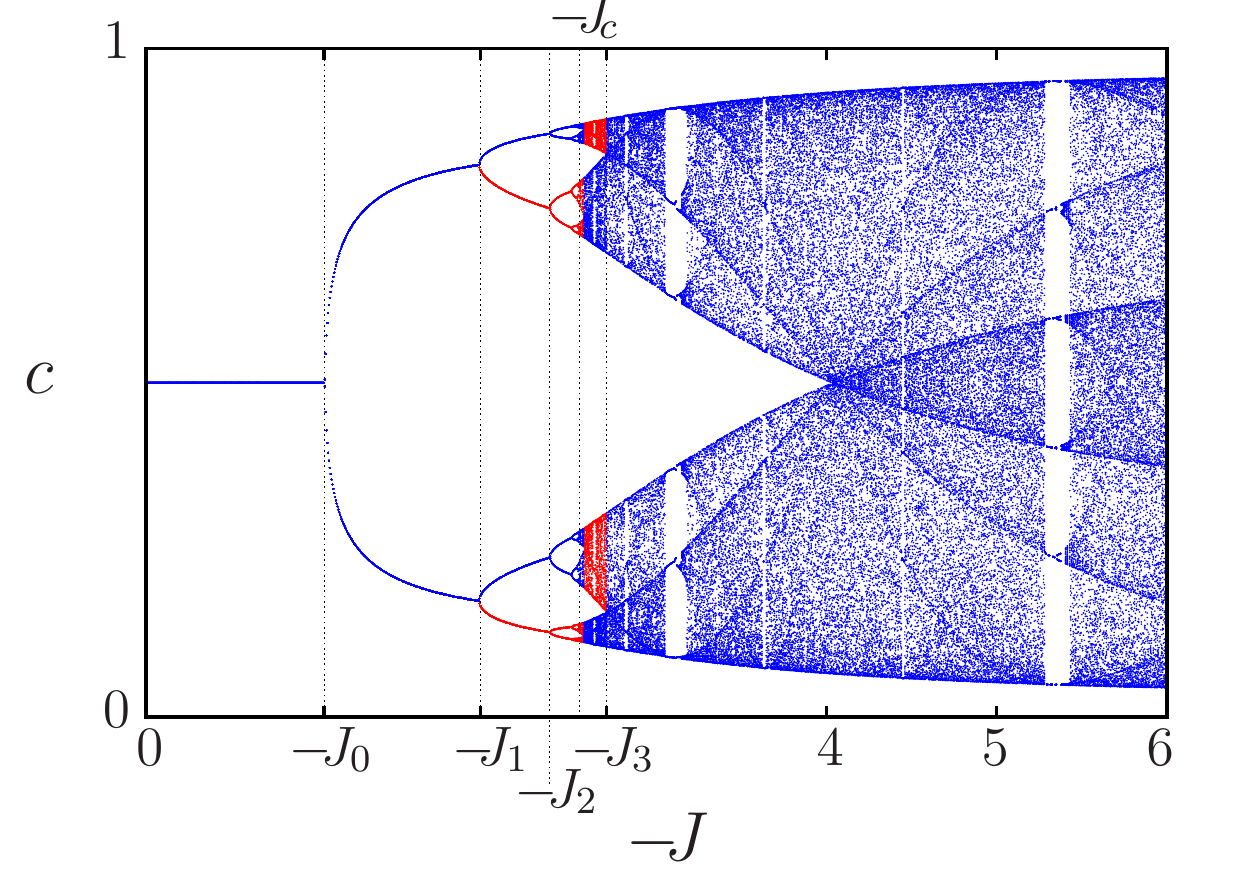}
  \includegraphics[width=0.48\columnwidth]{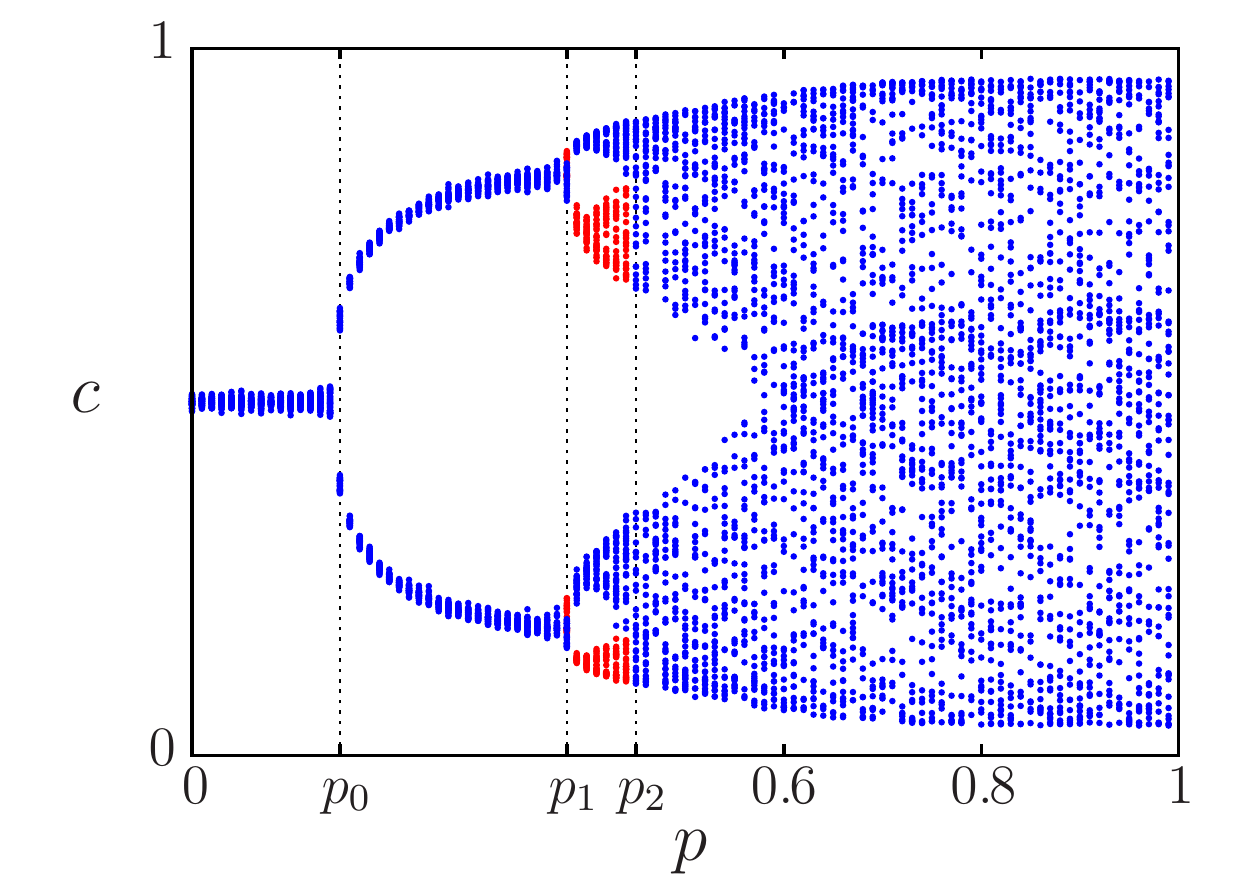}
 \caption{\label{fig:mfb} (left) Bifurcation diagram of
   the mean field map, Eq.~\eqref{eq:mf}, by varying $J$.  The
   doubling bifurcation route to chaos ends at $J=J_c$. For $0>J\geq
   J_2$ and $J_3>J\geq 6$ there is only one attractor (blue, darker dots). For
   $J_2>J\geq J_c$ there are two, one corresponding to the lower
   branches that bifurcate up to $J_c$ (red, lighter dots), and the other one
   to the upper branches (blue, darker dots).  For $J_c>J\geq J_3$ there are
   two chaotic attractors, one corresponding to the lower branches
   (blue, darker dots), the other to the top branches (red, lighter dots).  For every
   value of $J$, the dots are 64 iterates of the map after a transient of $10^3$ time steps. For
   values of $J$ with only one basin of attraction the orbits do not
   depend on the initial average opinion $c(t=0)$.  For values of $J$
   that correspond to two attractors, one of them was found with
   $c(0)=0.1$, the other one with $c(0)=0.9$.  (right) Small-world
    probabilistic bifurcation diagrams as functions of the long range
    probability $p$. For $p\lesssim p_0$ there are almost periodic orbits of period one
    and for $p_0\lesssim p\lesssim p_1$ of period two. For $p_1\lesssim p\lesssim p_2$
    we find two attractors, one (in red, lighter) in the lower branches,
    the other one (in blue, darker) in the top ones. }
 \end{center}
\end{figure}

The opinion of agent $i$ changes in time according to the transition
probability $\tau(s_i|h_i)$ that agent $i$ will hold the opinion $s_i$
at time $t+1$ given the local opinion $h_i$ at time $t$. This
transition probability, shown in Fig.~\ref{fig:mfeq}-left, is given by
\eq[tau]{
 \tau(h) =
  \begin{cases}
    \varepsilon & \text{if $h<q$,}\\
    \dfrac{1}{1+\exp(-2J(2h-1))} & \text{if $q\le h\le 1-q$,}\\
    1-\varepsilon & \text{if $h>1-q$,}
  \end{cases}
}
with $\tau(h)=\tau(1|h)$.

The simplest mean-field description of the model is given by 
\eq[eq:mf]{
  c' = f(c)=\sum_{w=0}^k \binom{k}{w} c^w (1-c)^{k-w} \tau\left(\dfrac{w}{k}\right),
}
with $c'=c(t+1)$ and $c=c(t)$.  The term in 
parenthesis on the {\em r.h.s} of this expression denotes the 
$w$-combinations from a set of $k$ elements. In Fig.~\ref{fig:mfeq}-right
we show some graphs of $f$. The bifurcation digram of this map after varying $J$ is shown in Fig.~\ref{fig:mfb}-left. 

\begin{figure}[t]
 \begin{center}
  \begin{tabular}{cc}
    \includegraphics[width=0.44\columnwidth]{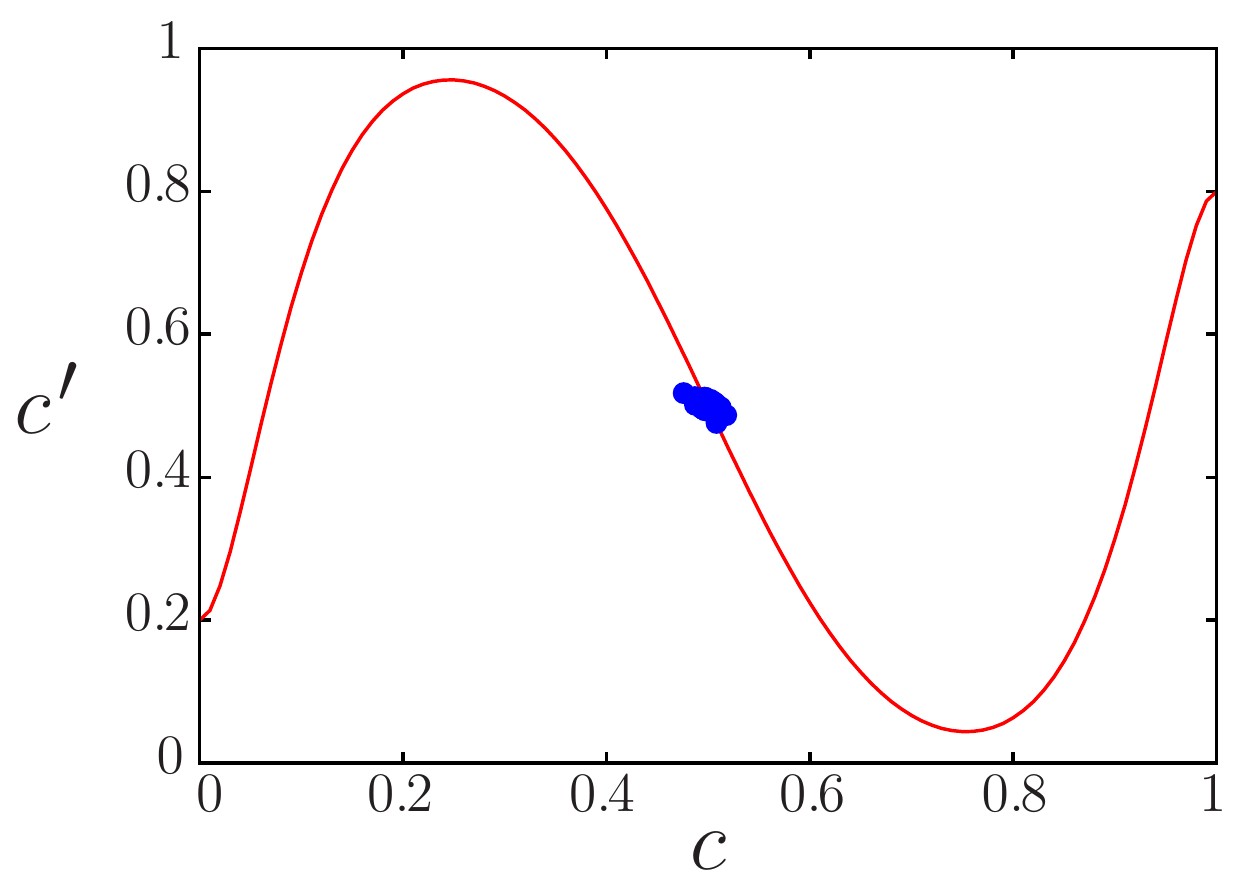}&
    \includegraphics[width=0.44\columnwidth]{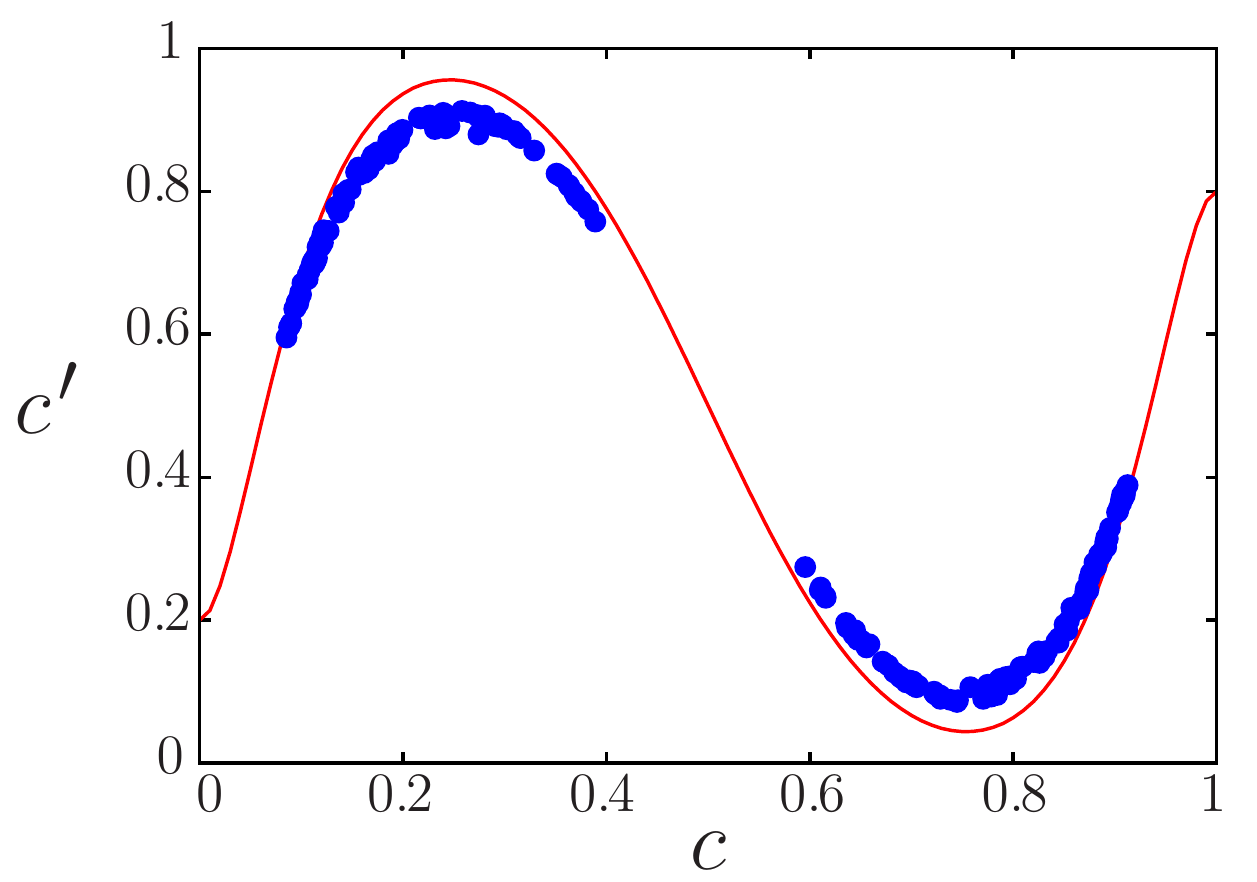} \\
    \includegraphics[width=0.44\columnwidth]{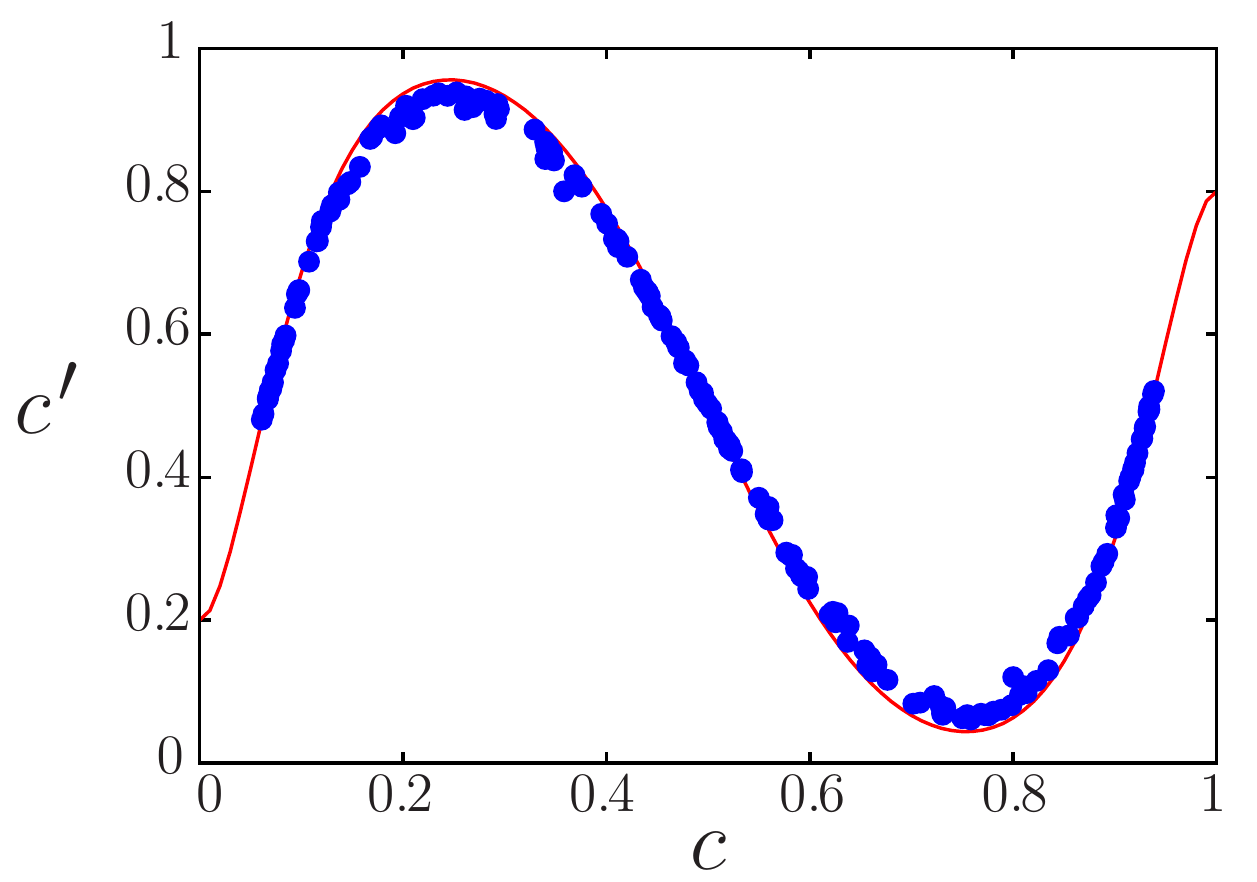} &
    \includegraphics[width=0.44\columnwidth]{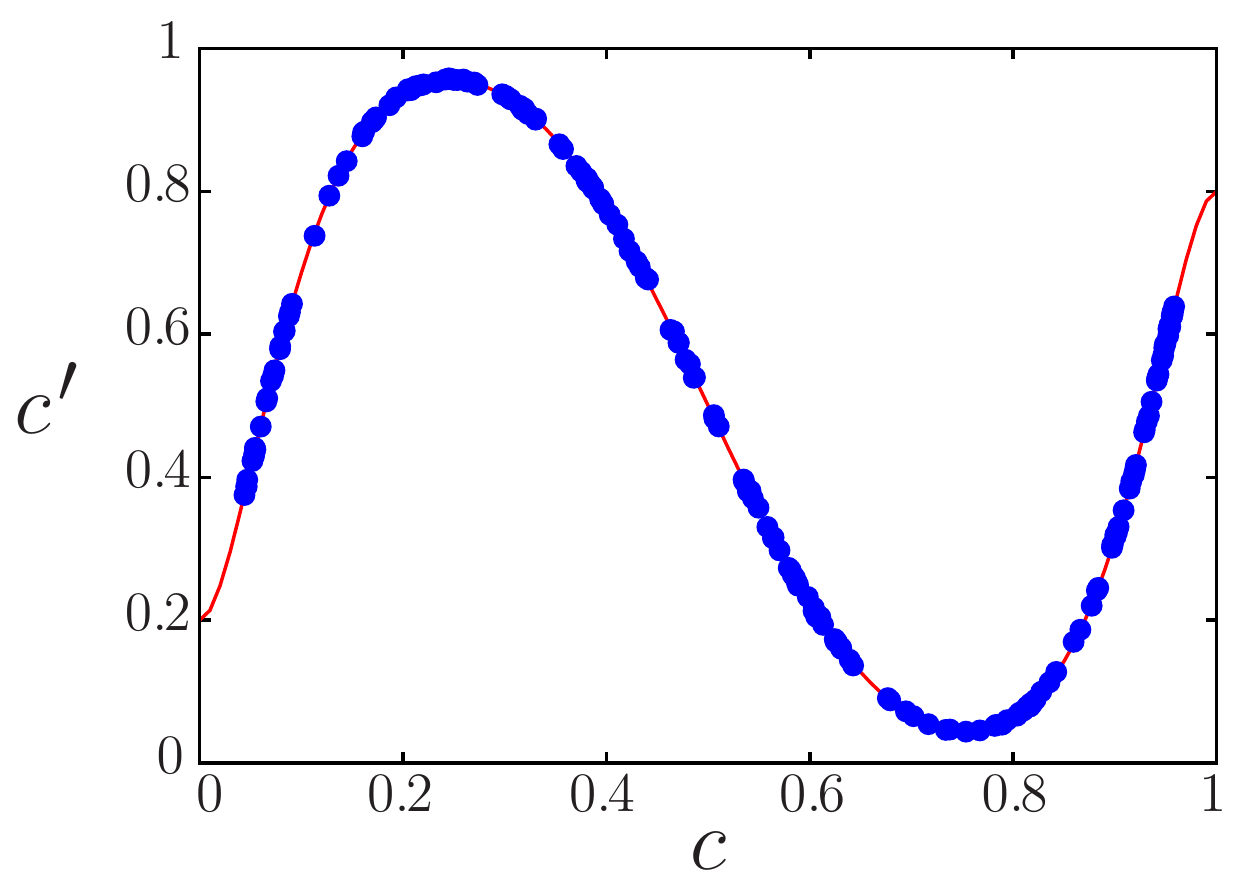} \\                                        
  \end{tabular}
  \end{center}
  \caption{\label{fig:conf} (Color online)  Return map of the
    average opinion $c$ on small-world networks for
    several values of the long-range connection probability $p$ with
    $J=-6$, $k=20$, $N=10^3$, and a transient
    of $10^3$ time steps. The following 200 iterations are shown as
    (blue, darker) dots. The (red, lighter) continuous curve is Eq.~\eqref{eq:mf}. From left to right, top to bottom:
    $p=0.0$, $p=0.5$,  $p=0.6$, and  $p=1.0$.}
\end{figure}

By varying the long-range probability $p$, we observe the transition towards the mean-field behaviour, as reported in Fig.~\ref{fig:conf}. This induces a stochastic bifurcation diagram by varying $p$, Fig.~\ref{fig:mfb}-right that is quite similar to that obtained in the mean-field approximation by varying $J$, Fig.~\ref{fig:mfb}-left. 

Notice that up to now we have met phase transition that, in the mean-field description, implies the change of stability of fixed points, while here we observe a real bifurcation diagram with coexistence of basins, period-doubling and chaos.

\subsection{Phenomenological renormalization group}

We can exploit the scaling form of the correlation function at the phase transition to obtain the phase boundary by means of renormalization technique. The idea is that of performing a sort of coarse-graining on a typical pattern, for instance by reducing block of $b$ spins or cells to just one spin, assigning to it the value of the majority in the block. In general, the resulting pattern will be typical of a different value of the control parameters (say $J$). The correlation length $\xi(J)$ has a value that depends on $J$ (more precisely: on the distance between $J$ and $J_c$,  the critical value at the transition, where $\xi$ diverges). After coarse graining, $\xi$ is reduced  by a factor $b$, so it will be typical of a value of $J$ farther from the transition. At the transition, however, $\xi$ diverges, so this value of the control parameter is an unstable fixed point of this procedure. With more than one parameter, one has one or more fixed points (that correspond to the various universality classes of the problem). The phase separation line is like a ridge between two valleys, while the saddle is the unstable fixed point. 

It is quite difficult to apply this procedure directly to patterns, but we can obtain a map directly for the parameters from the (somehow estimated) probability distributions. 
We illustrate it for the DK model and $b=2$~\cite{DKrenorm}.  
ddilution

The idea is the following: let us consider the joint probability $Q_1(x_1; y_1, y_2)$ of getting two site values $y_1, y_2$ at time $t$ and the value $x_1$ at  time $t+1$ 
\eq{
  Q_1(x_1; y_1, y_2) = \tau(x_1|y_1, y_2) \pi_2(y_1, y_2).
}

We can use this relation as a phenomenological definition of $\tau$ from observations,
\eq{
 \tau(x_1|y_1, y_2)  = \dfrac{Q(x_1, y_1, y_2)}{\pi_2(y_1, y_2)}. 
}

\begin{figure}[t]
 \begin{center}
   \includegraphics[width=0.78\columnwidth]{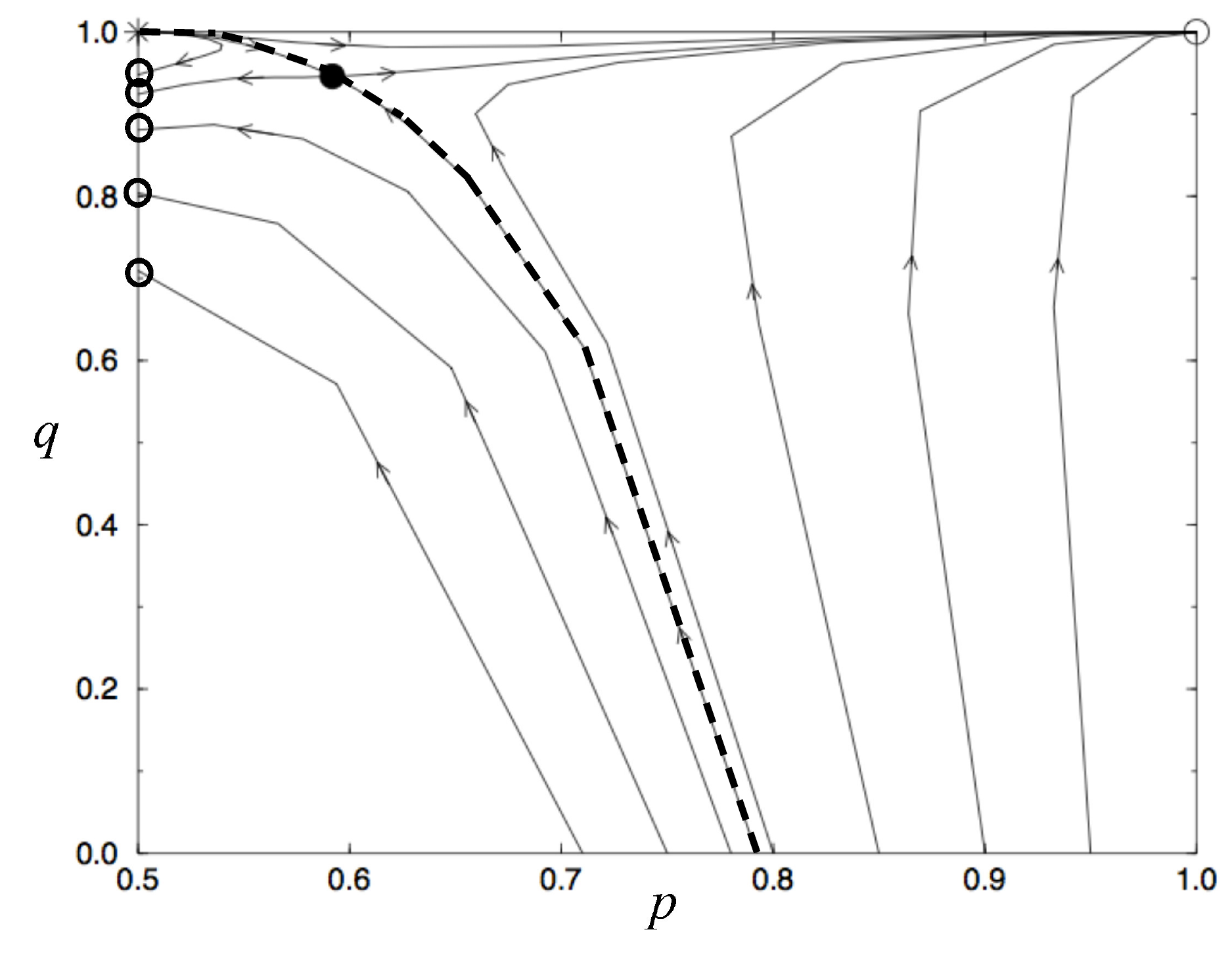}
 \caption{\label{fig:renorm} Phenomenological renormalization group for the Domany-Kinzel model with mean-field of order 4 (adapted from~Ref.~\cite{DKrenorm}). The trajectories that show a circle at $p=0.5$ finally end in the stable attracting point $p=q=0$. The other stable attracting point is $p=q=1$. The Ising $T=0$ point at $p=1/2, q=1$ is globally unstable. The saddle point marked with a filled dot determines the universality class (related to the eigenvalues of the Hessian in that point). The separatrix between the basins of the two stable points, marked with a dashed line, identifies the phase transition.  }
 \end{center}
 \end{figure}
The joint probability of larger structures, for instance $Q_2(x_1, x_2;z_1, z_2, z_3, z_4)$, is
\meq[eq:Q4]{
 Q_2(x_1,& x_2;y_1, y_2, y_3, y_4) =\\ 
  & \sum_{z_1, z_2, z_3}  \tau(x_1|z_1, z_2) \tau(x_2|z_2, z_3)\tau(z_1|y_1, y_2) \cdot \\
  &\qquad  \tau(z_2|y_2, y_3)\tau(z_3|y_3, y_4) \pi_4(y_1, y_2, y_3, y_4).
}
We  implement a block coarse graining procedure that keeps the symmetries of the model, in this case we want to stay on the $w=0$ plane (i.e., keeping the absorbing state). Thus the coarse-graining procedure is given by $R(0|0,0) = R(1|0,1)=R(1|1,0)=R(1|1,1)=1$ (with $\sum_x R(x|y,z)=1$). Applying it to $Q_2$ and to $\pi_4$ we get 
\meq{
 \tilde Q_2(\tilde x_1; \tilde y_1, \tilde y_2) &= \sum_{x_1, x_2}\sum_{y_1, y_2, y_3, y_4} R(\tilde x_1|x_1, x_2) \cdot \\
 & R(\tilde y_1|y_1,y_2) R(\tilde y_2|y_3,y_4) Q_2(x_1, x_2;y_1, y_2, y_3, y_4), 
}
\meq{
 \tilde \pi_4(\tilde y_1, \tilde y_2) &= \sum_{y_1, y_2, y_3, y_4} R(\tilde y_1|y_1,y_2) \cdot\\
 &\qquad  R(\tilde y_2|y_3,y_4) \pi_4(y_1, y_2, y_3, y_4), 
 }
and thus
\eq{
 \tilde Q_2(\tilde x_1; \tilde y_1, \tilde y_2) = \tilde \tau(\tilde x_1|\tilde y_1, \tilde y_2) \tilde \pi_4(\tilde y_1, \tilde y_2), 
}
from where we get get
\eq{
 \tilde \tau(\tilde x_1|\tilde y_1, \tilde y_2) = \dfrac{\tilde Q_2(\tilde x_1; \tilde z_1, \tilde z_2)}{ \tilde \pi_4(\tilde y_1, \tilde y_2)}. 
}
In this way we obtain the renormalization map from $\tau$ to $\tilde \tau$ shown in Fig.~\ref{fig:renorm}. 

The maps shown that in the plane $(p_1, p_)$ there are two fully attractive fixed points($(0, 0)$ and $(1, 1)$), one fully repulsive fixed point $(1/2, 1)$ and one nontrivial fixed point $(p^*_1, p^*_2)$. The attractive fixed point $(0, 0)$ is related to the absorbing state whereas the other attractive fixed point $(1,1)$ is related to the active state. Almost all trajectories are attracted to either one or the other of these two points as can be seen in figure 1. The base of attraction are separated by a line, which should be identified as the critical line of the Domany-Kinzel model. The separatrix hits the line $p_2 = 1$ at $p_1 = 1/2$, which is the fully repulsive fixed point (the compact percolation line).

In order to implement this procedure we need an estimation of the asymptotic probability distribution that can be obtained with the techniques of Section~\ref{sec:meanfield}. 
One can see from Fig.~\ref{fig:renorm}-right that even with a simple mean-field of order 4, one can accurately estimate the phase boundary.

\section{Conclusions}

We have illustrated some aspects of phase transitions in probabilistic cellular automata, trying to illustrate how such a problem  arises in different contexts and some of the method used for its study.

\section*{acknowledgement}
This work was partially supported by  EU projects 288021 (EINS -- Network of Excellence in Internet Science)  and project PAPIIT-DGAPA-UNAM IN109213.

\end{document}